\documentclass[aps, prb,superscriptaddress, twocolumn,preprintnumbers,amsmath,amssymb]{revtex4-1}
\usepackage{graphicx}
\usepackage{verbatim}
\usepackage{bm}
\usepackage{amssymb, latexsym, amsmath, textcomp}

\def\bx{\boldsymbol{x}}
\def\by{\boldsymbol{y}}

\def\br{\boldsymbol{r}}
\def\bR{\boldsymbol{R}}
\def\bk{\boldsymbol{k}}
\def\bK{\boldsymbol{K}}
\def\bq{\boldsymbol{q}}

\def\bS{\boldsymbol{S}}
\def\bG{\boldsymbol{G}}

\def\bsigma{\boldsymbol{\sigma}}

\def\rSU2{{\rm SU}(2)}

\def\bt{\boldsymbol{t}}

\begin{document}
\title{Majorana spin liquids and projective realization of ${\rm SU}(2)$ spin symmetry}
\author{Gang Chen}
\affiliation{Department of Physics, University of Colorado, Boulder, Colorado 80309, USA}
\author{Andrew Essin}
\affiliation{Department of Physics, University of Colorado, Boulder, Colorado 80309, USA}
\author{Michael Hermele}
\affiliation{Department of Physics, University of Colorado, Boulder, Colorado 80309, USA}

\date{\today}
\begin{abstract}
We revisit the fermionic parton approach to $S = 1/2$ quantum spin liquids with ${\rm SU}(2)$ spin rotation symmetry, and the associated projective symmetry group (PSG) classification.  We point out that the existing PSG classification is incomplete; upon completing it, we find spin liquid states with $S = 1$ and $S = 0$ Majorana fermion excitations coupled to a deconfined $Z_2$ gauge field.  The crucial observation leading us to this result is that, like space group and time reversal symmetries, spin rotations can act projectively on the fermionic partons;  that is, a spin rotation may be realized by simultaneous ${\rm SU}(2)$ spin and gauge rotations.  We show that there are only two realizations of spin rotations acting on fermionic partons:  the familiar naive realization where spin rotation is not accompanied by any gauge transformation, and a single type of projective realization.  We discuss the PSG classification for states with projective spin rotations.  To illustrate these results, we show that there are four such PSGs on the two-dimensional square lattice.  We study the properties of the corresponding states, finding that one -- with gapless Fermi points -- is a stable phase beyond mean-field theory.  In this phase, depending on parameters, a small Zeeman magnetic field can open a partial gap for the Majorana fermion excitations.  Moreover, there are nearby gapped phases supporting $Z_2$ vortex excitations obeying non-Abelian statistics.  We conclude with a discussion of various open issues, including the challenging question of where such $S = 1$ Majorana spin liquids may occur in models and in real systems.
\end{abstract}

\maketitle

\section{Introduction}
\label{sec:intro}

Some of the most intriguing states of matter are those beyond the conventional classification of phases according to spontaneously broken symmetry, band theory, and Fermi liquid theory.  The classic examples of such so-called exotic states are the fractional quantum Hall liquids,\cite{tsui82,laughlin83} which are characterized by topological order\cite{wen90} and associated properties such as fractionally charged excitations.\cite{laughlin83}  Quantum spin liquids\cite{anderson87} -- ground states of Mott insulators with no spontaneously broken symmetry -- are another class of states that are in many cases exotic, and are often also characterized by topological order and fractionalized excitations.\cite{sachdev04, senthil04, alet06, lee06,lee08,balents10}  A number of experiments over the last several years have uncovered materials where exotic quantum spin liquids may be present.\cite{lee08,balents10}

In systems with $S = 1/2$ local moments and ${\rm SU}(2)$ spin rotation symmetry, $S = 1/2$ spinons generally play an important role in the theory of spin liquid states, whether as quasiparticle excitations, as formal objects in terms of which the theory is constructed, or both.  Such spinons can be bosons or fermions, and can also obey fractional or non-Abelian statistics.  

Several recent works have raised the fascinating prospect of spin liquids where spinon quasiparticles do not carry $S = 1/2$, but are instead $S = 1$ Majorana fermions.\cite{wangf10, yao11, biswas11,lai11, lai11b}  In particular, such states were shown to occur in exactly solvable ${\rm SU}(2)$-invariant generalizations\cite{wangf10,yao11,lai11,lai11b} of Kitaev's  honeycomb lattice model.\cite{kitaev06} Of course, such models are rather special, and an approach to study $S = 1$ Majorana spin liquids in more general $S = 1/2$ spin models is a desirable complement to the exact solutions.
Biswas \emph{et. al.} have made a fascinating proposal in this direction,\cite{biswas11} constructing a mean-field theory of such a state on the triangular lattice.  We note that Majorana fermions also play an important role in other spin liquids of recent interest with ${\rm SU}(2)$ spin symmetry.\cite{greiter09, xu10a}

In this paper we show that, surprisingly, $S = 1$ Majorana spin liquids fit naturally into a well-known construction of spin liquids based on $S = 1/2$ fermionic partons.  Such spin liquid states can be classified in terms of their projective symmetry group (PSG).\cite{wen02}  We show that the existing 
PSG classification of $S = 1/2$ fermionic parton states is incomplete; upon completing it, we find the $S = 1$ Majorana spin liquids.  The spin liquids we find have, in addition to a triplet of $S = 1$ Majorana fermions, a spin singlet Majorana fermion.  Going beyond mean-field theory, all of these fermions are coupled to a deconfined $Z_2$ gauge field.  Moreover, wavefunctions for these states are easily obtained via Gutzwiller projection.

Before delving into details, let us first make the above assertions plausible.  In the $S = 1/2$ fermionic parton approach, space group and time reversal symmetries act projectively on fermions -- that is, such operations are realized as a product of the naive operation combined with an appropriate ${\rm SU}(2)$ gauge transformation.\cite{wen02}  The crucial observation, which to our knowledge has not been made before, is that spin rotations can also act projectively on the $S = 1/2$ partons.  We show that there are only two distinct realizations of spin rotation symmetry -- a familiar naive realization, where spin rotations alone are a symmetry, and a projective realization, where only combined spin and gauge rotations are a symmetry.  The projective realization can be thought of as a ``locking together'' of ${\rm SU}(2)$ spin symmetry and ${\rm SU}(2)$ gauge transformations, and is analogous to  color-flavor locking in high-density quantum chromodynamics.\cite{alford08}  

We find that $S = 1$ Majorana spin liquids occur when spin rotations are realized projectively.
The spin symmetry of such states is not readily apparent if one works in terms of $S = 1/2$ partons.  Manifest spin rotation invariance is recovered upon writing the $S = 1/2$ partons in terms of Majorana fermions that turn out to transform as singlets and triplets under projective spin rotations.

When spin rotations are realized projectively, the PSG classification needs to be re-done.  We show that Majorana spin liquid PSGs -- henceforth referred to as Majorana PSGs -- are in one-to-one correspondence with the ${\rm SU}(2)$ PSGs in the existing classification (\emph{i.e.} for the naive realization of spin rotation symmetry).  In general, for a given lattice there are only a few ${\rm SU}(2)$ PSGs, and therefore there are not many Majorana PSGs, at least if we insist on keeping time-reversal invariance and all lattice symmetries, as we do -- for simplicity -- in this paper.  For example, on the square lattice there are four Majorana PSGs,\cite{wen02} and on an anisotropic triangular lattice there are two.\cite{zhou02}  On the perfect (isotropic) triangular lattice we show that there are actually \emph{no} Majorana PSGs. This occurs because the mean-field Hamiltonian for a Majorana spin liquid breaks time reversal (and usually also reflection symmetry) if it has a non-bipartite structure of hoppings; therefore, unless one allows breaking of some symmetries, many frustrated lattices are not expected to admit any Majorana spin liquids.  On the other hand, if we broaden our scope to allow for breaking of time reversal and some lattice symmetries, Majorana spin liquids are certainly possible on frustrated lattices.  In Sec.~\ref{sec:discussion}, we speculate on the implications of these observations for finding Majorana spin liquids in realistic models and in experiments.

Beyond development of the general results mentioned above, we discuss Majorana spin liquids on the square lattice.  The primary purpose of this discussion is to give a concrete illustration of our more general results, and to discuss the properties of some spin liquid states arising from our construction.  There are four Majorana PSGs on the square lattice; for each of these we discuss the state with the simplest mean-field Hamiltonian (\emph{i.e.} with only the shortest-ranged hopping allowed by symmetry).  Three of these states have nested Majorana Fermi surfaces, and we expect these do not describe stable spin liquids beyond mean-field theory.  One state, however, is characterized by gapless Fermi points, and we show that it is a stable phase.  This state is dubbed the MB1-Dirac state.  The behavior of this state in a Zeeman magnetic field is interesting; depending on parameters, a small Zeeman field  opens a gap for some of the Majorana fermions.  Similar effects of Zeeman field were noted in the exactly solvable model of Ref.~\onlinecite{lai11}.  The Fermi points of the MB1-Dirac state acquire a full gap upon introducing either a weak columnar dimerization or a weak breaking of parity and time-reversal.  In the presence of both these orders, it is possible to have a state where $Z_2$ vortices are bound to an odd number of Majorana zero modes and carry non-Abelian statistics, as in the B-phase of the Kitaev honeycomb lattice model.\cite{kitaev06}

Our mean-field theory is distinct from that of Biswas \emph{et. al.} (Ref.~\onlinecite{biswas11}): in their formalism there is only a triplet of Majorana fermions, where in ours there is also a spin singlet Majorana fermion.  Understanding the relationship, if any, between these two mean-field theories is an open problem.  We contrast the two approaches in Appendix~\ref{app:bfls}.  Briefly, our formalism can incorporate fluctuations about mean-field theory using standard ideas of slave particle gauge theories, while Ref.~\onlinecite{biswas11} seems to require a more novel approach, which would be interesting to study in detail.  Projected wavefunctions can be easily obtained in our formalism, while it is not yet clear how to do this following Biswas \emph{et. al.}  These distinctions notwithstanding, it should be emphasized that the states obtained via these two approaches are similar in their physical properties, and in some cases the two mean-field theories may even describe two different limits of the same phase.

We now  outline the remainder of the paper.  In Sec.~\ref{sec:parton}, we briefly review the fermionic parton approach to $S = 1/2$ spin liquids, and the classification of such states by projective symmetry group.  A variety of useful notation is also introduced there.  In Sec.~\ref{sec:projective}, we discuss the possibility of projective spin rotation symmetry, and show that there are only two possible realizations of spin rotation symmetry in the fermionic parton approach, subject only to some minimal assumptions.  Next, in Sec.~\ref{sec:majorana}, we study the most general mean-field fermion Hamiltonian invariant under projective spin rotations, and show that its single-particle excitations are $S=1$ and $S=0$ Majorana fermions.  We also discuss the low-energy effective $Z_2$ gauge theory of Majorana spin liquids, and make some comments about their projected wavefunctions.  Section~\ref{sec:mpsg} is concerned with the classification of Majorana PSGs.  In Sec.~\ref{sec:mapping} we establish the one-to-one correspondence between Majorana PSGs and ${\rm SU}(2)$ PSGs in the existing PSG classification.  In Sec.~\ref{sec:sq-psgs} we enumerate the four Majorana PSGs on the square lattice, and in Sec.~\ref{sec:frustrated} we discuss frustrated mean-field ans\"{a}tze for Majorana spin liquids and time reversal symmetry breaking.  In Sec.~\ref{sec:examples}, we study Majorana spin liquids on the square lattice at the mean-field level, considering each of the four PSGs.  The properties of the stable MB1-Dirac state are considered in more detail in Sec.~\ref{sec:mb1-properties}.  In particular, we show that the MB1-Dirac state is a stable phase and discuss the effects of Zeeman magnetic field.  We also consider the properties of some nearby gapped phases, including some with non-Abelian statistics of $Z_2$ vortices.  Various technical details, as well as a discussion of the approach of Ref.~\onlinecite{biswas11}, are contained in the appendices.  The paper concludes with a discussion in Sec.~\ref{sec:discussion}.

\section{Review of fermionic parton approach to $S = 1/2$ spin liquids}
\label{sec:parton}

We consider a system of $S = 1/2$ spins placed on the sites $\br$ of some regular lattice, with Hamiltonian
\begin{equation}
\label{eqn:hamiltonian}
H = \sum_{(\br, \br')} J_{\br \br'} \bS_{\br} \cdot \bS_{\br'} \text{.}
\end{equation}
Here the sum is over distinct pairs of sites $(\br, \br')$; for later applications we take each such pair to be ordered according to some arbitrary convention.  The operator $\bS_{\br}$ generates rotations of the spin at site $\br$, and satisfies the commutation relations $[ S^i_{\br}, S^j_{\br'} ] = i \delta_{\br \br'} \epsilon^{i j k} S^k_{\br}$.  Our focus is on systems obeying full ${\rm SU}(2)$ spin rotation symmetry, as well as time reversal and space group symmetry. The precise form of the Hamiltonian will be less important for us than its symmetry, since we are primarily concerned with constructing and classifying possible states -- the much more difficult problem of finding specific models that realize the new states we identify will be left for future work.  Even so, it is worth noting that in most cases of interest the exchange couplings $J_{\br \br'}$ will be predominantly antiferromagnetic (positive), although it is not necessary for all the exchange couplings to be positive.  Moreover, additional multi-spin exchange terms (not written) can also be included in the Hamiltonian.

We are interested in constructing possible spin liquid ground states, which are simply ground states that preserve all of the microscopic symmetries of the original model.  (Occasionally we will also use a looser definition of spin liquid that only requires the preservation of translation and spin rotation symmetry.)  One major approach to constructing spin liquid states begins by rewriting the spin operator $\bS_{\br}$ as a bilinear of partons (bosons or fermions) and ends in the construction of wavefunctions, as well as associated low-energy effective field theories.  For reasons discussed at the end of Sec.~\ref{sec:projective}, in this paper we shall confine our attention to fermionic partons.  In the remainder of this section we review the fermionic parton approach, mostly following Ref.~\onlinecite{wen02}, before proceeding to our results in Sec.~\ref{sec:projective}.  Our intent is not to provide a complete review, but rather to remind the reader of the basic facts, emphasizing those aspects important for connecting to the remainder of the paper.  To this end, we use notation differing from most treatments in the literature.

The spin operator is written as
\begin{equation}
\label{eqn:partons}
\bS_{\br} = \frac{1}{2} f^\dagger_{\br \alpha} \bsigma_{\alpha \beta} f^{\vphantom\dagger}_{\br \beta} \text{,}
\end{equation}
where $f^\dagger_{\br \alpha}$ creates a spin-$1/2$ fermion of spin $\alpha = \uparrow, \downarrow$ at site $\br$.  The fermions obey canonical anticommutation relations, $\bsigma = (\sigma^1, \sigma^2, \sigma^3)$ is a vector of the $2 \times 2$ Pauli matrices, and summation over repeated indices is implied.  In order to have a faithful representation of the spin model, we must also impose the local constraint
\begin{equation}
\label{eqn:u1-constraint}
f^\dagger_{\br \alpha} f^{\vphantom\dagger}_{\br \alpha} = 1 \text{.}
\end{equation}

It is clear from Eqs.~(\ref{eqn:partons}) and~(\ref{eqn:u1-constraint}) that there is a local  redundancy under ${\rm U}(1)$ gauge transformations $f_{\br \alpha} \to e^{i \phi_{\br}} f_{\br \alpha}$.  In fact the full local redundancy is known to be ${\rm SU}(2)$.\cite{affleck88,dagotto88}  To expose this, we introduce the $2 \times 2$ matrix
\begin{equation}
F_{\br} = \left( \begin{array}{cc}
f_{\br \uparrow} & f^\dagger_{\br \downarrow} \\
f_{\br \downarrow} & - f^\dagger_{\br \uparrow} \end{array} \right) \text{.}
\end{equation}
The reason for introducing this matrix, rather than working with a two-component spinor as is more common, is that left-${\rm SU}(2)$ rotations of $F_{\br}$ are spin rotations, while right-${\rm SU}(2)$ rotations are gauge transformations.  Therefore, when we work with objects built from $F_{\br}$, transformation properties under both spin and gauge rotations are manifest.
The spin operator can be written
\begin{equation}
\bS_{\br} = - \frac{1}{4} \operatorname{tr} ( \bsigma F^{\vphantom\dagger}_{\br} F^\dagger_{\br} ) \text{,}
\end{equation}
which is manifestly invariant under the ${\rm SU}(2)$ gauge transformation $F_{\br} \to F_{\br} U_{\br}$, where $U_{\br} \in {\rm SU}(2)$.  Moreover, gauge transformations are generated by
\begin{equation}
\bG_{\br} = \frac{1}{4} \operatorname{tr} ( F^{\vphantom\dagger}_{\br} \bsigma F^\dagger_{\br} ) \text{.}
\end{equation}
The constraint Eq.~(\ref{eqn:u1-constraint}) can be expressed as $G^3_{\br} = 0$, and this automatically implies also $G^1_{\br} = G^2_{\br} = 0$.  Therefore, we have the local ${\rm SU}(2)$ constraint
\begin{equation}
\bG_{\br} = 0 \text{.}
\end{equation}
A fact that will be useful later on is that $\bS_{\br}$ and $\bG_{\br}$ form a complete set of fermion bilinears for the single site $\br$.  We have the commutation relations
\begin{eqnarray}
\left[ S^i_{\br}, S^j_{\br} \right] &=& i \epsilon^{i j k} S^k_{\br} \\
\left[ G^i_{\br}, G^j_{\br} \right] &=& i \epsilon^{i j k} G^k_{\br} \\
\left[ S^i_{\br}, G^j_{\br} \right] &=& 0 \text{.}
\end{eqnarray}

To proceed, one writes down a quadratic mean-field Hamiltonian in terms of the fermions.  This Hamiltonian can be obtained by decoupling the quartic spin exchange interaction, but this is not necessary and we will not frame our discussion in that language.  The most general quadratic Hamiltonian satisfying $[H_0 , S^i ] = 0$, where $S^i = \sum_{\br} S^i_{\br}$, is
\begin{eqnarray}
H_0 &=& \sum_{(\br, \br')} \Big[ i \chi^0_{\br \br'} \operatorname{tr} (F^{\vphantom\dagger}_{\br} F^\dagger_{\br'} )
+ \chi^i_{\br \br'} \operatorname{tr} (F^{\vphantom\dagger}_{\br} \sigma^i F^\dagger_{\br'} ) \Big]  \nonumber \\
&+& \sum_{\br} a_0^i(\br) G^i_{\br} \text{,} \label{eqn:standard-h0}
\end{eqnarray}
where $a^i_0(\br)$, $\chi^0_{\br \br'}$ and $\chi^i_{\br \br'}$ ($i = 1,2,3$) are real parameters.  A choice of these parameters is referred to as a mean-field ansatz.  

A spin liquid wavefunction is generated from this Hamiltonian by Gutzwiller projection:
\begin{equation}
| \psi \rangle = {\cal P} | \psi_0 \rangle \text{,}
\end{equation}
where $| \psi_0 \rangle$ is the ground state of $H_0$, and ${\cal P}$ implements Gutzwiller projection (\emph{i.e.} projection onto the subspace satisfying the constraint $\bG_{\br} = 0$).  More precisely, ${\cal P} = \prod_{\br} {\cal P}_{\br}$, where
\begin{equation}
{\cal P}_{\br} = \frac{4}{3} \bS^2_{\br} = \frac{4}{3} \big( \frac{3}{4} - \bG^2_{\br} \big) \text{.}
\end{equation}
From this form it is clear that $G^i_{\br} | \psi \rangle = 0$.

The form of $H_0$ guarantees that $| \psi \rangle$ is a spin singlet: $S^i | \psi_0 \rangle = 0$, and therefore $S^i | \psi \rangle = S^i {\cal P} | \psi_0 \rangle = {\cal P} S^i | \psi_0 \rangle = 0$.\footnote{Note that the projected wavefunction may in principle be magnetically ordered, or may break  other symmetries besides spin rotation.  One can check for this possibility by calculating the spin correlation function (or other appropriate correlation functions) after projection.}  For $| \psi \rangle$ to describe a spin liquid, it must also preserve time reversal and space group symmetries (or a subgroup of these symmetries, if we use a looser definition of spin liquid).  This occurs exactly when $H_0$ is invariant under  projective symmetry group (PSG) transformations.\cite{wen02}  Consider a space group operation $S : \br \to S(\br)$.  This operation acts on spin operators by
\begin{equation}
\label{eqn:spin-transf}
S : \bS_{\br} \to \bS_{S(\br)} \text{,}
\end{equation}
and we require this operation to leave $| \psi \rangle$ invariant (possibly up to multiplication by an overall phase).  Acting on fermion operators, we have the projective transformation
\begin{equation}
\label{eqn:s-proj-transf}
S : F_{\br} \to F_{S(\br)} U^S_{\br} \text{,}
\end{equation}
where $U^S_{\br} \in {\rm SU}(2)$ is an arbitrary gauge transformation, which does not affect the transformation of the gauge-invariant spin operators.  In order for $S$ to be a symmetry, we require that there exists some choice of $U^S_{\br}$ such that Eq.~(\ref{eqn:s-proj-transf}) leaves $H_0$ invariant.  Similarly, time reversal is implemented as an anti-unitary operation sending
\begin{equation}
\label{eqn:t-proj-transf}
{\cal T} : F_{\br} \to (i \sigma^2) F_{\br} U^{\cal T}_{\br} \text{,}
\end{equation}
where again $U^{{\cal T}}_{\br}$ must be chosen to leave $H_0$ invariant.  Spin rotations are realized by $F_{\br} \to U F_{\br}$ for $U \in {\rm SU}(2)$, and do not require any gauge transformation (but see Sec.~\ref{sec:projective}).  Moreover, $H_0$ is also invariant under a subgroup of pure gauge transformations, called the invariant gauge group (IGG).  The IGG can be ${\rm SU}(2)$, ${\rm U}(1)$, $Z_2$, or products of these groups.  There is thus some freedom in the choice of $U^S_{\br}$ and $U^{\cal T}_{\br}$, since these gauge transformations can always be multiplied by an element of the IGG.  This situation is expressed by writing
\begin{equation}
\text{SG} = \text{PSG} / \text{IGG} \text{,}
\end{equation}
where SG stands for the symmetry group of the spin model.

To summarize this discussion, and give a more precise definition of PSG, we say that a PSG is specified by the following set of transformations (and also products of these transformations):
\begin{eqnarray}
S : F_{\br} &\to& F_{S(\br)} U^S_{\br} \label{eqn:sg-psg-trans} \\
{\cal T} : F_{\br} &\to& (i \sigma^2) F_{\br} U^{\cal T}_{\br}  \\
\text{Spin rotation} : F_{\br} &\to& U F_{\br} \label{eqn:naive-srot} \\
\text{IGG} : F_{\br} &\to& F_{\br} U^{\alpha}_{\br} \text{.} \label{eqn:igg-psg-trans}
\end{eqnarray}
Here, $S$ runs over all space group operations, and $\alpha$ parametrizes the elements of the IGG.  We also require that there exists a (non-zero) ansatz invariant under these transformations.  
Moreover, we require that the ansatz is invariant \emph{only} under those pure gauge transformations in the IGG.  
Two collections [Eqs.~(\ref{eqn:sg-psg-trans} - \ref{eqn:igg-psg-trans})] of such transformations are equivalent, and are considered realizations of the same PSG, if they are gauge-equivalent.  (We do not require the ans\"{a}tze associated with two sets of transformation laws to be gauge-equivalent.)

PSGs can be classified following Ref.~\onlinecite{wen02}, where the classification was worked out for the square lattice for the cases $\text{IGG} = {\rm SU}(2), Z_2$, where there are only a finite number of PSGs.  The classification was also partially worked out for $\text{IGG} = {\rm U}(1)$, where there are an infinite number of PSGs.
PSGs are often referred to by type of IGG; for example, if $\text{IGG} = Z_2$, we say that the PSG is a $Z_2$ PSG.  It should be remarked that a very similar PSG classification exists in the bosonic parton approach.\cite{wangf06}
The main difference is that the gauge structure for bosonic partons is only ${\rm U}(1)$ and not ${\rm SU}(2)$.  

So far we described how to generate a wavefunction from a mean-field ansatz, but we have said nothing about how to arrive at a low-energy effective theory.  We  sketch a prescription here, which builds on ideas introduced in Ref.~\onlinecite{senthil00}, and has subsequently been used in many works.  Beyond the brief discussion here, we also illustrate this prescription in greater detail, via a concrete example in Sec.~\ref{sec:majorana}.  
One introduces a dynamical lattice gauge field with gauge group given by IGG, and couples it to the fermions.\footnote{In the case $\text{IGG} = {\rm U}(1)$, the gauge field needs to be compact.  This is so because for non-compact ${\rm U}(1)$ gauge fields, the magnetic flux is a ${\rm U}(1)$ globally conserved density, and therefore introducing a non-compact gauge field would introduce an extra global symmetry not present in the original spin model.  This issue only arises when $\text{IGG} = {\rm U}(1)$.}  The Gauss' law constraint is chosen so that the gauge theory reduces to a $S = 1/2$ Heisenberg model in the strong coupling limit.  Barring accidental fine-tuning -- which is anyway easily corrected -- the resulting low-energy theory has precisely the global symmetries of the microscopic spin model of interest.  Moreover it reduces to a spin model in the same universality class (\emph{i.e.} with $S = 1/2$ spins, short-range interactions, and the same symmetries), in the strong coupling limit.  Therefore it is expected to be a legitimate low-energy effective theory, in the sense that its phases and phase transitions occur for some spin model in the same universality class as the microscopic spin model of interest.  Furthermore, such gauge theories also arise naturally upon studying fluctuations about mean-field theory.\cite{wen91}

The spin liquid phase is the deconfined phase of the low-energy effective gauge theory.  It should be noted that when $\text{IGG} = {\rm U}(1), {\rm SU}(2)$, there may not be a stable deconfined phase; in that case the low-energy theory does not describe a stable spin liquid phase.  When $\text{IGG} = Z_2$, deconfinement of the gauge field is robust, as it is protected by the non-zero energy gap to $Z_2$ vortex excitations.  

It is important to keep in mind that the classification of PSGs is not the same as a classification of spin liquid phases.  For instance, there can be distinct spin liquids with the same PSG.  This occurs, for example, when some parameter of a mean-field ansatz can be tuned to transform a state with a fermion gap into a gapless state.  

Finally, we briefly comment on the relationship between the projected wavefunction and effective theory obtained from the same mean-field ansatz.  In our opinion, this issue is poorly understood and in need of more attention in future work.  The effective theory, by design, correctly captures the universal long-wavelength physics of a given spin liquid phase; there is no guarantee that the wavefunction does the same.  Indeed, in a number of cases there is compelling evidence that the long-wavelength properties of projected wavefunctions do not match the corresponding effective theory.\cite{paramekanti04, hermele08,tay11}  On the other hand, a class of wavefunctions for $Z_2$ spin liquids does have the same $Z_2$ topological order as the effective gauge theory.\cite{ivanov02, paramekanti05}  Moreover, the wavefunctions provide short-distance information -- energetic information, for example -- that is inaccessible using the effective theory approach. Therefore it would be desirable to better understand the circumstances under which projected wavefunctions capture the correct long-distance behavior of the low-energy effective theory, and, in other circumstances, to learn how the wavefunctions may be improved.

\section{Projective realization of ${\rm SU}(2)$ spin symmetry}
\label{sec:projective}

Here, within the framework of the $S=1/2$ fermionic parton approach reviewed above, we consider the possibility that spin rotations are realized projectively.  That is, we consider the possibility that the mean-field Hamiltonian is not invariant under the naive spin rotation Eq.~(\ref{eqn:naive-srot}), but is invariant when naive spin rotation is combined with an appropriate gauge transformation.  Because spin rotations are a continuous symmetry, the conditions on how it may be realized are quite restrictive.
We will show that, subject to minimal assumptions, there are only two distinct ways to realize spin rotation symmetry: the naive transformation of Eq.~(\ref{eqn:naive-srot}), and the projective spin rotation
\begin{equation}
\text{Spin rotation} : F_{\br} \to U F_{\br} U^\dagger \text{.} \label{eqn:proj-srot}
\end{equation}
When $\text{IGG} = {\rm SU}(2)$, Eq.~(\ref{eqn:naive-srot}) and Eq.~(\ref{eqn:proj-srot}) are not distinct.  However, in Sec.~\ref{sec:majorana} below, we will consider the most general mean-field Hamiltonian invariant under projective spin rotations and see that generically $\text{IGG} = Z_2$.

We assume spin rotations are generated by the Hermitian operators $T^i$, and that $[H_0 , T^i] = 0$, where $H_0$ is the mean-field Hamiltonian.  Note that we do not assume $H_0$ has the form given in Eq.~(\ref{eqn:standard-h0}).  It follows from Noether's theorem and the fact that the mean-field Hamiltonian is quadratic that  $T^i$ is a fermion bilinear.
We make the following further assumptions:
\begin{enumerate}
\item $T^i = \sum_{\br} T^i_{\br}$.
\item For a gauge-invariant state $| \psi \rangle$ (which satisfies $G^i_{\br} | \psi \rangle = 0$), $T^i | \psi \rangle = S^i | \psi \rangle$.
\item $[T^i, T^j] = i \epsilon^{i j k} T^k$.
\end{enumerate}
While it might conceivably be possible to relax some of these assumptions, we have not found a sensible way to do this, and we will not consider this possibility here.

Here and throughout this paper, we will restrict attention to mean-field ans\"{a}tze that fully connect the lattice.  This means that  any two sites $\br_1$ and $\br_2$ are joined by a  path of lattice bonds $(\br, \br')$ such that, for each bond in the path, a fermion bilinear coupling  $F_{\br}$ with $F_{\br'}$ appears in $H_0$ with nonzero coefficient.  One reason for this restriction is just simplicity.  A deeper reason is that, for a mean-field ansatz where the lattice breaks into two or more disconnected components, the PSG classification reduces to a separate PSG classification for each disconnected component, and the IGG will be a product of IGGs for each of the disconnected components.  Therefore the more basic problem is to classify PSGs (and construct corresponding spin liquid states) for fully connected mean-field ansatz.

Since $S^i_{\br}$ and $G^i_{\br}$ are a complete set of single-site fermion bilinears, the most general form of $T^i_{\br}$ satisfying assumption (2) is
\begin{equation}
T^i_{\br} = S^i_{\br} + M^{i j}_{\br} G^j_{\br} \text{,}
\end{equation}
where $M^{i j}_{\br}$ is an arbitrary real $3 \times 3$ matrix.  For assumption (3) to hold we must also have $[T^i_{\br}, T^j_{\br} ] = i \epsilon^{i j k} T^k_{\br}$; it is shown in Appendix~\ref{app:mmatrix} that this implies either $M_{\br} = 0$ or $M_{\br} \in {\rm SO}(3)$.  We can therefore make a gauge transformation so that on every site either $M_{\br} = 0$ or $M^{i j}_{\br} = \delta^{i j}$.  Suppose that on one site $\br$, $M_{\br} = 0$, while on another $\br'$, $M^{i j}_{\br'} = \delta^{i j}$.  In this case, there is no spin rotation invariant fermion bilinear coupling $F_{\br}$ and $F_{\br'}$.  Since we assume the ansatz is fully connected, this means we must either have $M_{\br} = 0$ everywhere, or $M^{i j}_{\br} = \delta^{i j}$ everywhere.  

We have therefore shown that the only two possibilities for the generator of spin rotations are $T^i = S^i$, or
\begin{equation}
\label{eqn:proj-t}
T^i = S^i + G^i \text{,}
\end{equation}
where
\begin{equation}
G^i = \sum_{\br} G^i_{\br} \text{.}
\end{equation}
This form generates the projective spin rotations of Eq.~(\ref{eqn:proj-srot}).  We shall now proceed to study mean-field Hamiltonians, and the corresponding spin liquid states, where spin rotations are realized projectively in this fashion.

At this point, it is natural to ask whether any analogous results hold for bosonic partons.  Obviously the ${\rm SU}(2)$ gauge structure of fermionic partons is the crucial element in the above discussion, because it allows for a natural association, expressed in Eq.~(\ref{eqn:proj-srot}), of a gauge rotation with a given spin rotation.  Bosonic partons have only a ${\rm U}(1)$ gauge structure, so we expect that spin rotations cannot be realized projectively with bosonic partons, as long as the spin symmetry is ${\rm SU}(2)$.  On the other hand, if the spin symmetry is only ${\rm U}(1)$, we expect that projective spin symmetry can be realized with bosonic partons.  Study of this possibility is left for future work.

\section{Majorana spin liquids}
\label{sec:majorana}

Here, we consider the most general mean-field Hamiltonian invariant under projective spin rotation symmetry generated by $T^i = S^i + G^i$, and show that its single-particle excitations are $S = 1$ and $S = 0$ Majorana fermions.  Moreover, we show that $\text{IGG} = Z_2$, and write the low-energy effective $Z_2$ gauge theory describing the spin liquid state.  We also make some comments on projected wavefunctions.  The classification of Majorana PSGs, and specific examples of Majorana spin liquids, are discussed in later sections.

The most general quadratic Hamiltonian invariant under projective spin rotation symmetry is
\begin{equation}
H_0 = \sum_{(\br, \br')} \Big[ i \chi^1_{\br \br'} \operatorname{tr} (F^{\vphantom\dagger}_{\br} F^\dagger_{\br'} )
+ i \chi^2_{\br \br'} \operatorname{tr} ( \sigma^i F^{\vphantom\dagger}_{\br} \sigma^i F^\dagger_{\br'} ) \Big] \text{.}
\label{eqn:majorana-h0-Fs}
\end{equation}
The sum is over ordered pairs of lattice sites $(\br, \br')$; the ordering of the pairs is fixed but arbitrary.
If $\chi^2_{\br \br'} = 0$ for all bonds, then we have $\text{IGG} = {\rm SU}(2)$, in which case there is no distinction between projective and naive spin rotations.  Therefore we always want to consider $\chi^2_{\br \br'} \neq 0$ for some bonds.  The ground state of $H_0$ is a singlet under projective spin rotations, that is $T^i | \psi_0 \rangle = 0$, and also $\langle \psi_0 | T^i_{\br} | \psi_0 \rangle = 0$.  This implies that the corresponding projected wavefunction $| \psi \rangle = {\cal P} | \psi_0 \rangle$ satisfies $S^i | \psi \rangle = 0$ and $\langle \psi | S^i_{\br} | \psi \rangle = 0$.

As an aside, it is interesting to express $H_0$ directly in terms of $f_{\br \alpha}$-fermions; it takes the form
\begin{eqnarray}
H_0 &=& \sum_{(\br, \br')} \Big[ i (\chi^1_{\br \br'} + \chi^2_{\br \br'} ) f^\dagger_{\br \uparrow} f^{\vphantom\dagger}_{\br' \uparrow} + i (\chi^1_{\br \br'} - \chi^2_{\br \br'} ) f^\dagger_{\br \downarrow} f^{\vphantom\dagger}_{\br' \downarrow}  \nonumber \\
 &-& 2 i \chi^2_{\br \br'} f^{\vphantom\dagger}_{\br \uparrow} f^{\vphantom\dagger}_{\br' \uparrow} +  \text{H.c.} \Big]
 \text{.} \label{eqn:majorana-h0-fs}
\end{eqnarray}
This Hamiltonian combines an imaginary spin-dependent hopping, with an imaginary pairing of the up-spin fermions only.

Returning to the main task at hand, we define Majorana fermions as follows:
\begin{equation}
F_{\br} = \frac{1}{2} \big[  i s_{\br} + \bsigma \cdot \bt_{\br} \big] \text{,}
\label{eqn:f-in-majoranas}
\end{equation}
or, equivalently,
\begin{eqnarray}
s_{\br} &=& - i ( f^{\vphantom\dagger}_{\br \uparrow} - f^\dagger_{\br \uparrow} ) \\
t^1_{\br} &=& f^{\vphantom\dagger}_{\br \downarrow} + f^{\dagger}_{\br \downarrow} \\
t^2_{\br} &=& - i (f^{\vphantom\dagger}_{\br \downarrow} - f^{\dagger}_{\br \downarrow} )  \\
t^3_{\br} &=& f^{\vphantom\dagger}_{\br \uparrow} + f^\dagger_{\br \uparrow} \text{.}
\end{eqnarray}
We note that the same mapping was recently employed in Ref.~\onlinecite{burnell11} to study Kitaev's honeycomb lattice model using $S = 1/2$ partons.
These objects satisfy the anticommutation relations
\begin{eqnarray}
\left\{ s_{\br} , s_{\br'} \right\} &=& 2 \delta_{\br \br'} \\
\left\{ t^i_{\br} , t^j_{\br'} \right\} &=&  2 \delta^{i j} \delta_{\br \br'} \\
\left\{ s_{\br} , t^i_{\br'} \right\} &=& 0 \text{.}
\end{eqnarray}
Moreover, from the form Eq.~(\ref{eqn:f-in-majoranas}) it is clear that $s_{\br}$ is a singlet under projective spin rotations, while $\bt_{\br}$ transforms as a vector.

Expressing $H_0$ in terms of Majorana fermions we have
\begin{equation}
\label{eqn:majorana-h0}
H_0 = \sum_{(\br, \br')} \Big[ i \chi^s_{\br \br'} s_{\br} s_{\br'} + i \chi^t_{\br \br'} \bt_{\br} \cdot \bt_{\br'} \Big] \text{,}
\end{equation}
where $\chi^s_{\br \br'} = \chi^1_{\br \br'} / 2 + 3 \chi^2_{\br \br'} / 2$ and $\chi^t_{\br \br'} = \chi^1_{\br \br'}/2 - \chi^2_{\br \br'} / 2$.  From this form is it clear that we simply have a theory of decoupled singlet and triplet Majorana fermions. Moreover, as long as $\chi^2_{\br \br'} \neq 0$, the $s$-fermions and $t$-fermions have different spectra.

We also have the expressions
\begin{eqnarray}
S^i_{\br} &=& - \frac{1}{4} ( i s_{\br} t^i_{\br} + \frac{i}{2} \epsilon^{i j k} t^j_{\br} t^k_{\br} ) 
\label{eqn:s-in-majoranas} \\
G^i_{\br} &=&   \frac{1}{4} ( i s_{\br} t^i_{\br} - \frac{i}{2} \epsilon^{i j k} t^j_{\br} t^k_{\br} ) \\
T^i_{\br} &=&  - \frac{i}{4} \epsilon^{i j k} t^j_{\br} t^k_{\br} \text{.} \label{eqn:t-in-majoranas}
\end{eqnarray}
We note that $G^i_{\br} = 0$ if and only if
\begin{equation}
 t^1_{\br} t^2_{\br} t^3_{\br} s_{\br} = 1 \text{.} \label{eqn:majorana-constraint}
\end{equation}
This is a $Z_2$ form of the gauge constraint because the operator $ t^1_{\br} t^2_{\br} t^3_{\br} s_{\br}$ has eigenvalues $\pm 1$, and is the constraint appearing in the solution of Kitaev's honeycomb lattice model.\cite{kitaev06}

In Appendix~\ref{app:igg}, we show that $\text{IGG} = Z_2$.  This is done by directly proving that the only ${\rm SU}(2)$ gauge transformations leaving $H_0$ invariant are $U_{\br} = 1$ for all $\br$, and $U_{\br} = - 1$ for all $\br$.  It is also shown in Appendix~\ref{app:igg} that space group and time reversal operations leaving $H_0$ invariant must be of the form
\begin{eqnarray}
S : F_{\br} &\to& F_{S(\br)} \pi^S_{\br} \\
{\cal T} : F_{\br} &\to& (i \sigma^2) F_{\br} (i \sigma^2) \pi^{\cal T}_{\br}  \text{,} \label{eqn:general-form-of-T}
\end{eqnarray}
where $\pi^S_{\br}$ and  $\pi^{\cal T}_{\br}$ take values $\pm 1$ as a function of lattice site $\br$.

Now, we shall write down the low-energy effective $Z_2$ gauge theory describing a Majorana spin liquid state.  We note that essentially the same construction has been used previously for other $Z_2$ spin liquid states.\cite{senthil00}  For simplicity of notation, we consider an ansatz with only nearest-neighbor bonds.  On every nearest-neighbor bond we place an Ising degree of freedom with a two-dimensional Hilbert space, acted on by Pauli matrices $\sigma^z_{\br \br'} \equiv \sigma^z_{\br' \br}$ and $\sigma^x_{\br \br'} \equiv \sigma^x_{\br' \br}$.  $\sigma^z$ is the $Z_2$ vector potential, and $\sigma^x$ is the $Z_2$ electric field.  The local constraint becomes
\begin{equation}
 t^1_{\br} t^2_{\br} t^3_{\br} s_{\br} = \prod_{\br' \text{n.n.} \br} \sigma^x_{\br \br'} \text{,}
 \end{equation}
 where the product is over sites $\br'$ that are nearest-neighbors of $\br'$.  The Hamiltonian is
 \begin{eqnarray}
H &=& \sum_{\langle \br \br' \rangle} \sigma^z_{\br \br'} \Big[ i \chi^s_{\br \br'} s_{\br} s_{\br'} + i \chi^t_{\br \br'} \bt_{\br} \cdot \bt_{\br'} \Big] 
\nonumber \\
&-& h \sum_{\langle \br \br' \rangle} \sigma^x_{\br \br'}
- K \sum_{p} \prod_{\br \br' \in p} \sigma^z_{\br \br'} \text{.} \label{eqn:effective-h}
\end{eqnarray}
We take $h, K > 0$.  The sums in the first two terms are over nearest-neighbor bonds.  In the last term, the sum is over lattice plaquettes labeled by $p$, and the product $\prod_{\br \br' \in p}$ is a product over bonds in the perimeter of $p$.

It is highly nontrivial to obtain the ground state phase diagram of $H$; there may well be a variety of phases.  However, the physics is simple when $K$ is sufficiently large, which is where the spin liquid phase arises.  
In this limit the $Z_2$ gauge field enters its deconfined phase, where fluctuations of the $Z_2$ magnetic field $\prod_{\br \br' \in p} \sigma^z_{\br \br'}$ are suppressed.  The deconfinement is a robust property associated with a gap to $Z_2$ vortex excitations, which are plaquettes where $\prod_{\br \br' \in p} \sigma^z_{\br \br'} = -1$.  The other important quasiparticle excitations are the fermions themselves, which carry the $Z_2$ electric charge as evident from their minimal coupling to the $Z_2$ gauge field in Eq.~(\ref{eqn:effective-h}).  As for any state with a deconfined $Z_2$ gauge field, this state is characterized in part by its $Z_2$ topological order, as discussed for example in Ref.~\onlinecite{senthil01}.

Another important limit of $H$ arises when $h$ dominates over the other parameters.  To be concrete, we set $K = 0$ and assume $h \gg |\chi^s|, |\chi^t|$.  First setting $\chi^s = \chi^t = 0$, there is an extensively degenerate manifold of ground states, consisting of all states satisfying $\sigma^x_{\br \br'} = 1$ and the local constraint $ t^1_{\br} t^2_{\br} t^3_{\br} s_{\br} = 1$.  This is precisely the Hilbert space of the microscopic spin model.  The degeneracy can be resolved using standard degenerate perturbation theory in $\chi^s$ and $\chi^t$.  The first non-vanishing contribution occurs at second order, resulting in the effective Hamiltonian
\begin{equation}
H_{{\rm eff}} = \sum_{\langle \br \br' \rangle} J_{\br \br'} \bS_{\br} \cdot \bS_{\br'} \text{,}
\end{equation}
where
\begin{equation}
J_{\br \br'} =  \frac{4  \chi^t_{\br \br'} ( \chi^t_{\br \br'} + \chi^s_{\br \br'})}{h}   \text{.}
\end{equation}
It is interesting to note that $J_{\br \br'}$ is only antiferromagnetic when $\chi^t_{\br \br'} ( \chi^t_{\br \br'} + \chi^s_{\br \br'}) > 0$.  This suggests that states with negative $\chi^t_{\br \br'} ( \chi^t_{\br \br'} + \chi^s_{\br \br'}) < 0$ are not likely to occur in microscopic Hamiltonians dominated by antiferromagnetic exchange (but might reasonably occur if multi-spin exchanges are dominant).

We close this section with some brief comments on projected wavefunctions for Majorana spin liquids, which may be obtained by applying the usual Gutzwiller projection operator ${\cal P}$ to the ground state $| \psi_0 \rangle$ of $H_0$.  To simplify the discussion, we consider a state where $\chi^s = \chi^s_{\br \br'}$ and $\chi^t = \chi^t_{\br \br'}$ are non-zero only for $\br$ and $\br'$ nearest neighbors.  When $\chi^s = \chi^t$, then $\text{IGG} = {\rm SU}(2)$, and the projected wavefunction $| \psi \rangle = {\cal P} | \psi_0 \rangle$ can be associated with a low-energy effective  ${\rm SU}(2)$ gauge theory, keeping in mind the caveats mentioned at the end of Sec.~\ref{sec:parton}.  Now, if $\chi^s / \chi^t$ is changed continuously from unity, the wavefunction $| \psi \rangle$ does not change at all, because the pre-projected ground state $| \psi_0 \rangle$ does not change.  This occurs because $|\psi_0\rangle$ is a product of $s$-fermion and $t$-fermion ground state wavefunctions.  It is interesting to note that this is so even though the IGG of the mean-field state -- and hence the gauge group of the low-energy effective theory -- is now $Z_2$, as long as $\chi^s / \chi^t \neq 1$.  This is a rather dramatic illustration of the problematic association between low-energy effective gauge theories and projected wavefunctions (see Sec.~\ref{sec:parton}).

In order to obtain a distinct Majorana spin liquid projected wavefunction, we need to change $H_0$ in such a way that  $|\psi_0\rangle$ becomes different from the $\chi^s = \chi^t$ ground state.  One possibility is to change the \emph{sign} of $\chi^s / \chi^t$, as $|\psi_0 \rangle$ does change when $\chi^s / \chi^t$ crosses through zero.  This  only gives a single new wavefunction, as $|\psi_0\rangle$ is the same for all $\chi^s / \chi^t < 0$.  Another possibility is, rather than simply varying the ratio $\chi^s / \chi^t$, to add further-neighbor hopping for, say, the $s$-fermions and not for the $t$-fermions.  It will be interesting to study such wavefunctions in future work.

\section{Majorana Spin Liquid Projective Symmetry Groups}
\label{sec:mpsg}

Here, we show that, on any lattice, there is a one-to-one mapping between $Z_2$ PSGs for Majorana spin liquids (Majorana PSGs), and ${\rm SU}(2)$ PSGs.  Since classification of ${\rm SU}(2)$ PSGs has already been done for some lattices, we can exploit those results to give a classification of Majorana PSGs.  In this section, we first establish the mapping between Majorana PSGs and ${\rm SU}(2)$ PSGs (Sec.~\ref{sec:mapping}).  Next, we enumerate the four Majorana PSGs on the square lattice (Sec.~\ref{sec:sq-psgs}).  Finally, we consider frustrated mean-field ans\"{a}tze of the form Eq.~(\ref{eqn:majorana-h0}).  We say an ansatz is frustrated if it contains at least one closed loop with an odd number of bonds, so that $\chi^s_{\br \br'}$ (or $\chi^t_{\br \br'}$) is nonzero for each bond in the loop.  For instance, ans\"{a}tze with triangular plaquettes are frustrated.
In Sec.~\ref{sec:frustrated}, we explain that any frustrated ansatz breaks time-reversal symmetry, and also note that there are no time-reversal-symmetric Majorana PSGs on the isotropic triangular lattice.
 
\subsection{Correspondence with ${\rm SU}(2)$ projective symmetry groups}
\label{sec:mapping}

We now show that, on any lattice, there is a one-to-one correspondence between Majorana PSGs and ${\rm SU}(2)$ PSGs.  Precisely, we prove the following two statements:  (1) Given a Majorana PSG, there is a corresponding unique ${\rm SU}(2)$ PSG, where the symmetry operations are realized exactly as in the Majorana PSG.  (2) Given a ${\rm SU}(2)$ PSG, one can transform to a gauge where the symmetry operations are realized exactly as they are in a unique corresponding Majorana PSG.  In fact, the gauge needed in (2) is precisely the gauge used to classify SU(2) PSGs in Ref.~\onlinecite{wen02}.

We begin by showing (1).   A Majorana PSG is completely specified by the following collection of transformations under space group operations $S$, time reversal ${\cal T}$, spin rotations, and $Z_2$ IGG operations:
\begin{eqnarray}
S: F_{\br} &\to& \pi^S_{\br} F_{S(\br)} \label{eqn:maj-sg-op} \\
{\cal T}: F_{\br} &\to& \pi^{{\cal T}}_{\br} (i \sigma^2) F_{\br} (i \sigma^2) \label{eqn:maj-time-rev} \\
\text{Spin rotation} : F_{\br} &\to& U F_{\br} U^\dagger \\
\text{IGG} : F_{\br} &\to& \pm F_{\br}  \text{.}
\end{eqnarray}
Here, $\pi^S_{\br}, \pi^{\cal T}_{\br} = \pm 1$ take values $\pm 1$ as a function of $\br$.
There is also an ansatz of the form Eq.~(\ref{eqn:majorana-h0-Fs}) invariant under these transformations.  It is shown in Appendix~\ref{app:igg} that the forms Eq.~(\ref{eqn:maj-sg-op}) and Eq.~(\ref{eqn:maj-time-rev}) are the most general forms possible for an ansatz of the form Eq.~(\ref{eqn:majorana-h0-Fs}).  Such a set of transformation laws can be mapped into an equivalent set (\emph{i.e.} same PSG) under a gauge transformations of the $Z_2$ form $F_{\br} \to \pi_{\br} F_{\br}$, where $\pi_{\br}$ takes values $\pm 1$ as a function of $\br$.

Now, we are free to continuously change $\chi^1_{\br \br'}$ and $\chi^2_{\br \br'}$, as long as we change them in the same fashion on symmetry-related bonds.  In particular we can continuously tune $\chi^2_{\br \br'}$ to zero on all bonds, thus obtaining an ansatz with ${\rm SU}(2)$ IGG.  This ansatz is clearly invariant under the same set of PSG transformations given above, and we have thus shown (1).

Next, we show statement (2).  Suppose we have a ${\rm SU}(2)$ PSG [with spin rotation symmetry realized naively according to Eq.~(\ref{eqn:naive-srot})], where the IGG is generated by
\begin{equation}
\tilde{G}^i = \sum_{\br} M^{i j}_{\br} G^j_{\br} \text{.}
\end{equation}
Since $\tilde{G}^i$ must obey a ${\rm SU}(2)$ Lie algebra, we know from Appendix~\ref{app:mmatrix} that either $M_{\br} \in {\rm SO}(3)$ or $M_{\br} = 0$.  We can  make a gauge transformation so that  $M_{\br}$ is either the identity matrix or zero.  In fact, none of the $M_{\br}$ can be zero.  Suppose $M_{\br} = 0$ for some site $\br$, and is nonzero for some other site $\br'$.  In this case, there is no fermion bilinear joining $\br$ to $\br'$ that is invariant under the IGG.  Since we want to consider only fully connected ans\"{a}tze, we must then have $M_{\br} = 1$ on all sites.  ($M = 0$ on all sites is also possible, but this would mean the IGG is not ${\rm SU}(2)$.)
Since $M_{\br} \in {\rm SO}(3)$ for all $\br$, the gauge transformation needed to turn $\tilde{G}^i$ into $G^i$ is unique, up to multiplication by an arbitrary gauge transformation in the $Z_2$ center of ${\rm SU}(2)$.  Upon completing the mapping to a Majorana PSG, this $Z_2$ freedom will correspond to the $Z_2$ gauge freedom to map one set of Majorana PSG transformations into another equivalent set.

We have now gone to a gauge where $[ H_0 , G^i ] =0$, where $G^i = \sum_{\br} G^i_{\br}$.  By assumption, we also have $[H_0, S^i ] = 0$.  The most general quadratic Hamiltonian with these symmetries is
\begin{equation}
\label{eqn:su2_h0}
H_0 = i \sum_{(\br, \br')} \chi_{\br \br'} \operatorname{tr} (F^{\vphantom\dagger}_{\br} F^\dagger_{\br'} ) \text{.}
\end{equation}

Next, consider a space group operation $S$, which acts by
\begin{equation}
S : F_{\br} \to F_{S(\br)} U^S_{\br} \text{.}
\end{equation}
Because we can multiply this transformation by any element of the IGG, we are free to choose $U^S_{\br_0} = 1$ for some arbitrary site $\br_0$.  Now consider a site $\br'_0$, joined to $\br_0$ by the bond $(\br_0, \br'_0)$.  The gauge transformation $U^S_{\br}$ must transform the Hamiltonian on this bond into another bond Hamiltonian of the same form [as given in Eq.~(\ref{eqn:su2_h0})].  This only happens if $U^S_{\br'_0} = \pm 1$.  This conclusion  holds for the whole lattice, because by assumption we can connect any site $\br$ to $\br_0$ by some path of nonzero bonds.  Therefore, we have
\begin{equation}
S : F_{\br} \to \pi^S_{\br} F_{S(\br)} \text{,}
\end{equation}
where $\pi^S_{\br}$ takes values $\pm 1$ as a function of $\br$.

We proceed in essentially the same fashion for time reversal, which acts by
\begin{equation}
{\cal T} : F_{\br} \to (i \sigma^2) F_{\br} U^{{\cal T}}_{\br} \text{.}
\end{equation}
For an arbitrary site $\br_0$, we choose $U^{{\cal T}}_{\br_0} = ( i \sigma^2)$.  By the same argument as above, on all other sites we then have $U^{{\cal T}}_{\br} = \pm (i \sigma^2)$, leading to the desired result
\begin{equation}
{\cal T} : F_{\br} \to \pi^{{\cal T}}_{\br} (i \sigma^2) F_{\br} (i \sigma^2) \text{,}
\end{equation}
where $\pi^{\cal T}_{\br}$ takes values $\pm 1$ as a function of $\br$.  

Finally, we note that by making appropriate IGG transformations, we can choose spin rotations to act in the projective form $F_{\br} \to U F_{\br} U^\dagger$.  We have thus transformed to a gauge where the symmetry operations specify a unique Majorana PSG, and have shown statement (2).

\subsection{Majorana projective symmetry groups on the square lattice}
\label{sec:sq-psgs}

The square lattice space group is generated by the following operations:
\begin{eqnarray}
T_x : (r_x, r_y) &\to& (r_x + 1 , r_y) \\
T_y : (r_x, r_y) &\to& (r_x , r_y + 1) \\
P_x : (r_x, r_y) &\to& (-r_x, r_y) \\
P_{xy} : (r_x, r_y) &\to& (r_y, r_x) \text{.}
\end{eqnarray}
To specify a Majorana PSG, it is enough to specify the action of these operations, as well as time reversal, on the fermion operators.

Exploiting the mapping between ${\rm SU}(2)$ and Majorana PSGs, and exploiting the results on classification of ${\rm SU}(2)$ PSGs on the square lattice in Ref.~\onlinecite{wen02}, we find there are four Majorana PSGs on the square lattice.  We call these MA1, MA2, MB1 and MB2.  Ref.~\onlinecite{wen02} refers to the corresponding ${\rm SU}(2)$ PSGs as ${\rm SU}(2)$A$n0$, ${\rm SU}(2)$A$0n$, ${\rm SU}(2)$B$n0$, ${\rm SU}(2)$B$0n$, respectively.  To specify each of these PSGs, it is enough to give the action of the symmetry action on the singlet $s_{\br}$ fermions.  The triplet $t^i_{\br}$ obey identical transformation laws.  In Sec.~\ref{sec:examples}, we give example mean-field states obeying each of these four PSGs.

The MA1 PSG is specified by:
\begin{eqnarray}
T_x : s_{r_x, r_y} &\to& s_{r_x+1, r_y} \\
T_y : s_{r_x, r_y} &\to& s_{r_x, r_y+1} \\
P_x : s_{r_x, r_y} &\to& (-1)^{r_x} s_{-r_x, r_y} \\
P_{xy} : s_{r_x, r_y} &\to& s_{r_y, r_x} \\
{\cal T} : s_{r_x, r_y} &\to& (-1)^{(r_x + r_y)} s_{r_x, r_y} \text{.}
\end{eqnarray}

The MA2 PSG is specified by:
\begin{eqnarray}
T_x : s_{r_x, r_y} &\to& s_{r_x+1, r_y} \\
T_y : s_{r_x, r_y} &\to& s_{r_x, r_y+1} \\
P_x : s_{r_x, r_y} &\to& (-1)^{r_y} s_{-r_x, r_y} \\
P_{xy} : s_{r_x, r_y} &\to& s_{r_y, r_x} \\
{\cal T} : s_{r_x, r_y} &\to& (-1)^{(r_x + r_y)} s_{r_x, r_y} \text{.}
\end{eqnarray}

The MB1 PSG is specified by:
\begin{eqnarray}
T_x : s_{r_x, r_y} &\to& s_{r_x+1, r_y} \label{eqn:mb1-psg-first} \\
T_y : s_{r_x, r_y} &\to& (-1)^{r_x} s_{r_x, r_y+1} \\
P_x : s_{r_x, r_y} &\to& (-1)^{r_x} s_{-r_x, r_y} \\
P_{xy} : s_{r_x, r_y} &\to& (-1)^{r_x r_y} s_{r_y, r_x} \\
{\cal T} : s_{r_x, r_y} &\to& (-1)^{(r_x + r_y)} s_{r_x, r_y} \text{.} \label{eqn:mb1-psg-last}
\end{eqnarray}

The MB2 PSG is specified by:
\begin{eqnarray}
T_x : s_{r_x, r_y} &\to& s_{r_x+1, r_y} \\
T_y : s_{r_x, r_y} &\to& (-1)^{r_x} s_{r_x, r_y+1} \\
P_x : s_{r_x, r_y} &\to& (-1)^{r_y} s_{-r_x, r_y} \\
P_{xy} : s_{r_x, r_y} &\to& (-1)^{r_x r_y} s_{r_y, r_x} \\
{\cal T} : s_{r_x, r_y} &\to& (-1)^{(r_x + r_y)} s_{r_x, r_y} \text{.}
\end{eqnarray}

\subsection{Frustrated ans\"{a}tze and time-reversal symmetry breaking}
\label{sec:frustrated}

It is worth noting that, for all the square lattice Majorana PSGs, the form of time reversal constrains the hopping to have a bipartite structure.  That is, $\chi^s_{\br \br'}$ and $\chi^t_{\br \br'}$ can only be nonzero if one of $\br, \br'$ lies in the A sublattice, and the other site lies in the B sublattice.  Therefore, on the square lattice, any frustrated mean-field ansatz of the form Eq.~(\ref{eqn:majorana-h0}) breaks time reversal symmetry.  [We say an ansatz is frustrated if it contains at least one closed loop with an odd number of bonds, so that $\chi^s_{\br \br'}$ (or $\chi^t_{\br \br'}$) is nonzero for each edge in the loop.]

In fact, this statement holds on any lattice for Majorana spin liquid ans\"{a}tze.  Suppose we have a frustrated ansatz where $\chi^s_{\br \br'} \neq 0$ on the bonds of an odd-length closed loop.  (We could just as well consider a loop with $\chi^t_{\br \br'} \neq 0$.)  The time reversal operation takes the general form given in Eq.~(\ref{eqn:general-form-of-T}); on the $s_{\br}$ fermions we have
\begin{equation}
{\cal T} : s_{\br} \to \pi^{\cal T}_{\br} s_{\br} \text{.}
\end{equation}
If we choose $\pi^{\cal T}_{\br} = 1$, this transformation simply sends $\chi^s_{\br \br'} \to - \chi^s_{\br \br'}$.  For loops of even length this change can be compensated by an appropriate choice of gauge transformation $\pi^{\cal T}_{\br}$, but for a loop of odd length this is impossible, so any frustrated ansatz breaks time reversal symmetry.

To illustrate the strong restriction this imposes on Majorana spin liquid ans\"{a}tze, we consider the isotropic triangular lattice with 6-fold rotation symmetry and full translation symmetry of the triangular lattice.  Using only these symmetries, it is easy to show that any ansatz contains a closed loop of length three (this is true even if the ansatz has vanishing nearest-neighbor hopping).  Therefore, if one insists on time reversal invariance, there are no Majorana PSGs on the isotropic triangular lattice.

\section{Majorana spin liquids on the square lattice}
\label{sec:examples}

In this section we study the simplest ansatz for each of the four Majorana PSGs on the square lattice.  The ans\"{a}tze we consider are the simplest in the sense that they  include only the shortest-distance hopping permitted by symmetry.  In the case of MA1 and MB1 PSGs, this is nearest-neighbor hopping, while for MA2 and MB2 PSGs, we have fourth-neighbor hopping.  Each state is associated with a corresponding mean-field ${\rm SU}(2)$ spin liquid, obtained by setting $\chi^s = \chi^t$.  For the MA1 state this is the uniform resonating valence bond (RVB) state, and for the MB1 state this is the $\pi$-flux state.  

For each state, we study the mean-field excitation spectrum.  In all cases except the MB1 state, we find a nested Fermi surface, which we expect to give rise to instabilities upon going beyond mean-field theory and including short-range interactions of fermions.  However, at least for the MA1 and MB2 states, these instabilities are expected to be logarithmic in nature (similar to BCS instability), and therefore may play a role only at very low temperatures.  The MA2 state has discrete points with a $z=4$ excitation spectrum (\emph{i.e.} energy goes like the fourth power of momentum), which are likely to lead to a stronger instability that will be important at higher temperatures.  The likely consequence of these  instabilities is magnetic order, but it should be noted that the resulting ordered states are exotic, supporting gapped $Z_2$ vortex excitations and gapped fermionic spinons.  This occurs because deconfinment of the $Z_2$ gauge field is protected by the $Z_2$ vortex gap, and is thus robust to arbitrary small perturbations.

In contrast to the other states, the MB1 state has gapless Dirac points, and is thus dubbed the MB1-Dirac state.
The MB1-Dirac state is stable to arbitrary small perturbations, provided that space group, time reversal, and spin rotation symmetries are respected.  This is shown in Sec.~\ref{sec:mb1-properties}, where some other properties of the MB1-Dirac state are also discussed.

\subsection{MA1 state}
\label{sec:ma1}

The simplest mean-field ansatz with MA1 PSG has the Hamiltonian
\begin{equation}
H_0 = i \chi^s \sum_{\br} ( s_{\br} s_{\br + \bx} + s_{\br} s_{\br + \by} )  + i \chi^t \sum_{\br} ( \bt_{\br}  \cdot \bt_{\br + \bx} + \bt_{\br}  \cdot \bt_{\br + \by} ) \text{.}
\end{equation}
When $\chi^s = \chi^t$, this Hamiltonian reduces to pure imaginary hopping of the $S = 1/2$ $f_{\br \alpha}$ fermions, with the hopping phases such that there is zero magnetic flux through each plaquette.  Therefore, we obtain the ${\rm SU}(2)$ uniform RVB state at this special point.

Since both $s$- and $t$-fermions have the same spectrum, only with different coefficients, it is enough to focus on $H_{0 s}$, the $s$-fermion part of $H_0$.  Diagonalizing $H_{0s}$, we find the single-particle spectrum is given by
\begin{equation}
E_s(\bk) = 4 | \chi_s | \big| \sin k_x + \sin k_y \big| \text{,}
\end{equation}
where $-\pi \leq k_x, k_y \leq \pi$, with the proviso that $\bk$ and $-\bk$ are equivalent points due to the Majorana nature of the fermions.  

There are lines of gapless excitations for $k_y = - k_x$ and $k_y = \pm \pi + k_x$, which constitute the familiar diamond-shaped Fermi surface of the uniform RVB state.  Moreover, it is straightforward to show that invariance under MA1 PSG transformations requires these lines to be gapless at the quadratic level.  This means, without breaking some symmetry, it is impossible to remove the Fermi surface nesting by adding further neighbor hopping.
Upon going beyond mean-field theory and incorporating short-range fermion interactions, it is natural to expect that the nested nature of the Fermi surface leads to an instability to Neel magnetic order.  This is well-known to occur when $\chi^s = \chi^t$, and we have verified it more generally, treating an on-site interaction in mean-field theory, and finding an instability to Neel order for arbitrarily small interaction.\cite{unpublished_mft}

\subsection{MA2 state}
\label{sec:ma2}

\begin{figure}
\includegraphics[width=2in]{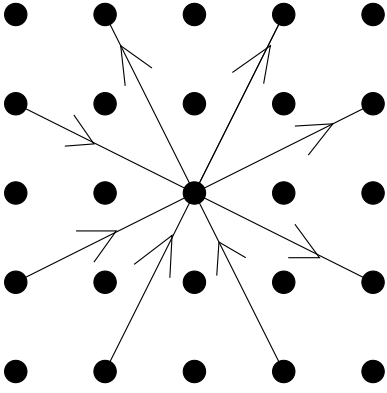}
\caption{Orientation of fourth-neighbor hoppings $\chi^s_{\br \br'}$ in the MA2 state.  All other hopping amplitudes can be obtained by a translation of those shown.}
\label{fig:ma2}
\end{figure}

The simplest mean-field ansatz with MA2 PSG has the following singlet part of the Hamiltonian:
\begin{eqnarray}
H_{0s} = H_{0s} &=& i \chi_s \sum_{\br} \Big[ s_{\br} s_{\br + 2 \bx + \by} + s_{\br} s_{\br + \bx + 2 \by}  \nonumber \\
&+& s_{\br} s_{\br - \bx + 2 \by} - s_{\br} s_{\br - 2 \bx + \by} \Big] \text{.}
\end{eqnarray}
For this PSG, first, second and third neighbor hoppings are required to vanish, but fourth-neighbor hopping is allowed by symmetry.  It is important to remember that hopping of Majorana fermions carries an orientation; the pattern of orientations for the fourth-neighbor hopping of $H_{0s}$ is illustrated in Fig.~\ref{fig:ma2}.  The single-particle spectrum is given by
\begin{eqnarray}
E_s(\bk) &=& 4 |\chi_s| \Big| \sin(2 k_x + k_y) + \sin(k_x + 2 k_y) \nonumber \\
&-& \sin(k_x - 2 k_y) + \sin(2 k_x - k_y) \Big| \text{,}
\end{eqnarray}
where $-\pi \leq k_x, k_y \leq \pi$ (again remembering that $\bk$ and $-\bk$ are equivalent points).  There are lines of gapless excitations for $k_y = -k_x$, $k_y = \pm \pi + k_x$, $k_y = \pm \pi/2$, and $k_x = \pm \pi/2$.  It can be shown that these gapless lines are protected (at the quadratic level) by  MA2 PSG transformations.  Moreover, we again expect an instability to magnetic order due to Fermi surface nesting.

It is interesting to note that the spectrum at $\bk = (\pi/2, -\pi/2)$ has a $z = 4$ character.  Expanding to lowest order near this point, we have
\begin{equation}
E_s[ \bk - (\pi/2, -\pi/2) ] \sim 8 | \chi^s | | k_x k^3_y - k^3_x k_y | \text{.}
\end{equation}
While we have not studied the issue in detail, the strong infrared singularities from such a point may lead to a strong instability to magnetic order (stronger than the usual logarithmic instability arising from Fermi surface nesting).  We also note that we have not verified whether the $z=4$ nature of this point is protected by MA2 PSG transformations.

\subsection{MB1 state}
\label{sec:mb1}

\begin{figure}
\includegraphics[width=2in]{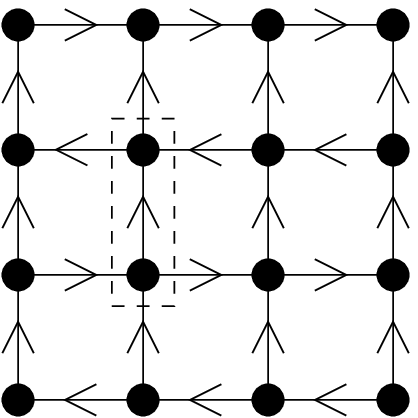}
\caption{Orientation of nearest-neighbor hoppings in the MB1 state.  The two-site unit cell is indicated by the dashed line.  Unit cells are labeled by $\bR = n_x \bx + 2 n_y \by$.  Sites are labeled by the pair $(\bR, i)$, where $i = 1$ corresponds to the lower site, and the upper site is $i = 2$.}
\label{fig:mb1}
\end{figure}

The simplest mean-field ansatz with MB1 PSG has the following singlet part of the Hamiltonian:
\begin{eqnarray}
H_{0s} &=& i \chi^s \sum_{\bR} \Big[ s_{\bR 1} s_{\bR 2} + s_{\bR 1} s_{\bR + \bx, 1}  \nonumber \\ 
&+& s_{\bR 2} s_{\bR + 2 \by, 1}
- s_{\bR 2} s_{\bR + \bx, 2} \Big]   \text{.}
\end{eqnarray}
We use a two-site unit cell as shown in Fig.~\ref{fig:mb1}, where the orientations of the nearest-neighbor hopping are also shown.  Sites are labeled by pairs $(\bR, i)$, as described in the caption of Fig.~\ref{fig:mb1}.  When $\chi^s = \chi^t$, this Hamiltonian reduces to pure imaginary hopping of the $S = 1/2$ $f_{\br \alpha}$ fermions, with the hopping phases such that there is magnetic flux of $\pi$ through each plaquette.  Therefore, we obtain the ${\rm SU}(2)$ $\pi$-flux state at this special point.

The single-particle spectrum is two-fold degenerate, and is given by
\begin{equation}
E_s(\bk) = 2 | \chi^s | \sqrt{4 - 2 \cos(2 k_x) - 2 \cos(2 k_y) } \text{,}
\end{equation}
where $-\pi \leq k_x \leq \pi$, $-\pi/2 \leq k_y \leq \pi/2$, and again it should be remembered that $\bk$ and $-\bk$ are equivalent points.  There are gapless Dirac nodes at $\bk = (0,0)$ and $\bk = (\pi,0)$, so we dub this state the MB1-Dirac state.  A low-energy theory for this state is derived in Appendix~\ref{app:mb1}, and physical properties of the state are discussed in Sec.~\ref{sec:mb1-properties}.  In particular, it is shown that the MB1-Dirac state is a stable spin liquid phase.

\subsection{MB2 state}
\label{sec:mb2}

\begin{figure}
\includegraphics[width=2in]{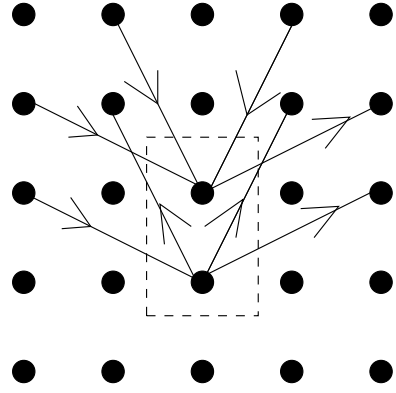}
\caption{Orientation of fourth-neighbor hoppings in the MB2 state.  The two-site unit cell is indicated by the dashed line.  Unit cells are labeled by $\bR = n_x \bx + 2 n_y \by$.  Sites are labeled by the pair $(\bR, i)$, where $i = 1$ corresponds to the lower site, and the upper site is $i = 2$.}
\label{fig:mb2}
\end{figure}

The simplest mean-field ansatz with MB2 PSG has the following singlet part of the Hamiltonian:
\begin{eqnarray}
H_{0s} &=& i \chi_s \sum_{\bR} \Big[ s_{\bR 1} s_{\bR + 2 \bx, 2} + s_{\bR 1} s_{\bR + \bx + 2 \by, 1}
+ s_{\bR 1} s_{\bR - \bx + 2 \by, 1}  \nonumber \\ 
&-& s_{\bR 1} s_{\bR - 2 \bx, 2} 
+  s_{\bR 2} s_{\bR + 2 \bx + 2 \by, 1} - s_{\bR 2} s_{\bR + \bx + 2 \by, 2}  \nonumber \\
&-& s_{\bR 2} s_{\bR - \bx + 2 \by, 2} - s_{\bR 2} s_{\bR - 2 \bx + 2 \by, 1} \Big] \text{.}
\end{eqnarray}
We again use a two-site unit cell as shown in Fig.~\ref{fig:mb2}, where the orientations of the fourth-neighbor hopping are also shown.  As in the MA2 state, first, second and third neighbor hopping is forbidden by symmetry, but fourth-neighbor hopping is allowed.

The single-particle spectrum is two-fold degenerate, and is given by
\begin{eqnarray}
E_s(\bk) &=& 8 \sqrt{2} | \chi_s | \big| \cos(k_x) \cos(k_y) \big| \nonumber \\
&\times&  \sqrt{2 - \cos(2 k_x) - \cos(2 k_y) } \text{,}
\end{eqnarray}
where $-\pi \leq k_x \leq \pi$, $-\pi/2 \leq k_y \leq \pi/2$, and again it should be remembered that $\bk$ and $-\bk$ are equivalent points.  There are gapless Dirac nodes at $\bk = (0,0)$ and $\bk = (\pi,0)$, as well as gapless lines for $k_x = \pm \pi / 2$.  It can be shown that the gapless points and lines are protected (at the quadratic level) by MB2 PSG transformations.  We again expect an instability to magnetic order due to Fermi surface nesting.

\section{Properties of MB1-Dirac state}
\label{sec:mb1-properties}

Here, we briefly discuss the low-energy effective theory of the MB1-Dirac state, and some of its physical properties.  In particular, we show that the MB1-Dirac state is a stable phase, and discuss the effect of Zeeman magnetic field.  We also discuss some nearby gapped symmetry-breaking phases, and show that states where $Z_2$ vortices are bound to an odd number of Majorana fermions (and thus have non-Abelian statistics) can occur when ${\cal P T}$-breaking order coexists with columnar dimer order.

The continuum low-energy theory (at the mean-field level) is worked out in Appendix~\ref{app:mb1}.  This theory is obtained by linearizing about the two gapless Dirac nodes at $\bk = (0,0)$ and $\bk = (\pi,0)$.  The imaginary-time action $S = \int d\tau d^2 \br {\cal L}$ is specified by the Lagrangian density
\begin{equation}
{\cal L} = \bar{\Psi} [ i \gamma_{\mu} \partial^s_{\mu} ] \Psi + 
\bar{\Phi}^i [ i \gamma_{\mu} \partial^t_{\mu} ] \Phi^i \text{.}
\end{equation}
Here, $\Psi$ is a four-component real fermion field arising from the singlet $s$-fermion on the lattice.  For each triplet fermion $t^i$, $\Phi^i$ is the corresponding four-component continuum field.  We denote the $2 \times 2$ Pauli matrices acting in the two-dimensional space of each node by $\tau^i$, and $\mu^i$ Pauli matrices act in the flavor space mixing the two nodes (see Appendix~\ref{app:mb1} for more detail).  The space-time index $\mu = 0,1,2$, and we define $\gamma_{\mu} = (\tau^2, \tau^1, -\tau^3)$.  The singlet (triplet) fermions have velocity $v_s$ ($v_t$), which enter via the derivatives $\partial^s_{\mu} \equiv (\partial_0 , v_s \partial_1, v_s \partial_2)$, and $\partial^t_{\mu} \equiv (\partial_0 , v_t \partial_1, v_t \partial_2)$.  Finally, we define $\bar{\Psi} = \Psi^T (-i \tau^2)$, and similarly for $\bar{\Phi}^i$.

This low-energy theory is in fact stable to the addition of small perturbations beyond mean-field theory, and the MB1-Dirac state is thus a stable phase.  Because excitations of the deconfined $Z_2$ gauge field are gapped, coupling to the $Z_2$ gauge field has no effect on the fermions at low energies and can be safely ignored.  While the resulting low-energy theory does not capture the $Z_2$ topological order that is present, it does correctly describe the universal behavior of correlation functions of local observables.  Moreover, it should be noted that ignoring fermion-gauge field coupling means  that our low-energy theory  -- if we add perturbations large enough to destabilize the MB1-Dirac state -- is only capable of describing phases where the $Z_2$ gauge field remains deconfined.

We also need to consider perturbations involving only the fermion fields.  The action $S$ is invariant under the renormalization group (RG) scale transformation $\tau \to e^\ell \tau$, $\br \to e^{\ell} \br$, $\Psi \to e^{-\ell} \Psi$, and $\Phi^i \to e^{- \ell} \Phi^i$.  This amounts to the statement that we have an RG fixed point with dynamic critical exponent $z = 1$, where the fermion fields have unit scaling dimension.  Perturbations that are relevant or marginal under this RG transformation can destabilize the phase; such perturbations are fermion bilinears with no derivatives (dimension two -- relevant), and with one derivative (dimension three -- marginal).  Making use of the symmetry transformations given in Appendix~\ref{app:mb1}, it can be shown that all bilinears with no derivatives are forbidden by the combination of spin rotation, space group and time reversal symmetries.  The only single-derivative bilinears allowed by symmetry are shifts of the velocities $v_s$ and $v_t$.  All other perturbations, including quartic interactions of fermions, are irrelevant under the RG.  Therefore the MB1-Dirac state is a stable phase.

\begin{figure}
\includegraphics[width=2.5in]{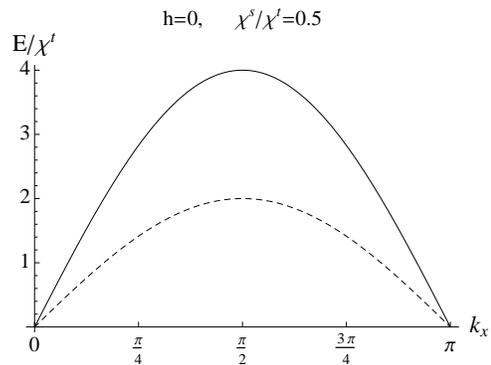}
\caption{Plot of the $h=0$ single-particle excitation spectrum for the MB1-Dirac state, along the line $k_y=0$, from $k_x = 0$ to $k_x = \pi$, for $\chi^s / \chi^t = 0.5$.  The solid line shows the six-fold degenerate $t$-fermion energy, while the dashed line is the two-fold degenerate $s$-fermion energy.  The gapless Dirac nodes are evident at $k_x = 0$ and $k_x = \pi$.}
\label{fig:zerob}
\end{figure}

\begin{figure}
\includegraphics[width=2.5in]{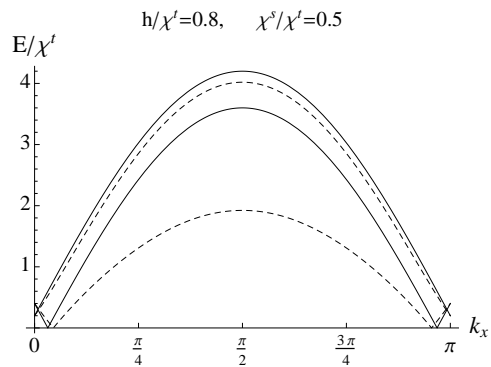}
\caption{Single-particle excitation spectrum of the MB1-Dirac state, for $h/\chi^t = 0.8$ and $\chi^s / \chi^t = 0.5$, along the line $k_y = 0$.  Here, the solid lines are the energies in the $t^1$-$t^2$ sector, and the dashed lines the energies in the $s$-$t^3$ sector.  Around the nodes, small Fermi pockets are opened in both sectors.  Away from the nodes, the $t$-fermion spectrum is split into three distinct branches.}
\label{fig:cspos}
\end{figure}

The same scaling considerations described above also imply the heat capacity $C(T) \propto T^2$ and the magnetic susceptibility $\chi(T) \propto T$.  Correlation functions of fermion bilinears fall off as $1/\br^4$ in space and $1/t^4$ in time.  This implies in particular that  $k = (0,0), (\pi,0), (0,\pi)$ and $(\pi,\pi)$ spin correlations fall off with these power laws.

\begin{figure}
\includegraphics[width=2.5in]{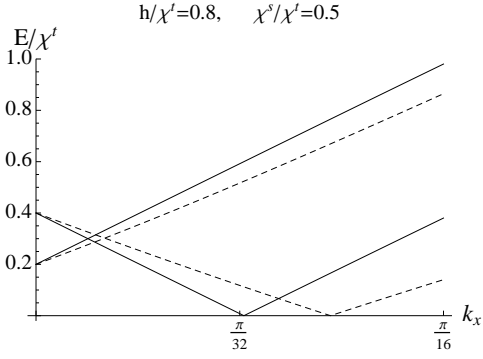}
\caption{Detail of the single-particle excitation spectrum of the MB1-Dirac state, along the line $k_y = 0$, near the node at $k_x = 0$.   The solid lines are the energies in the $t^1$-$t^2$ sector, and the dashed lines the energies in the $s$-$t^3$ sector. The parameters are the same as in Fig.~\ref{fig:cspos}, that is $h/\chi^t = 0.8$ and $\chi^s / \chi^t = 0.5$.}
\label{fig:cspos_small}
\end{figure}

\begin{figure}
\includegraphics[width=2.5in]{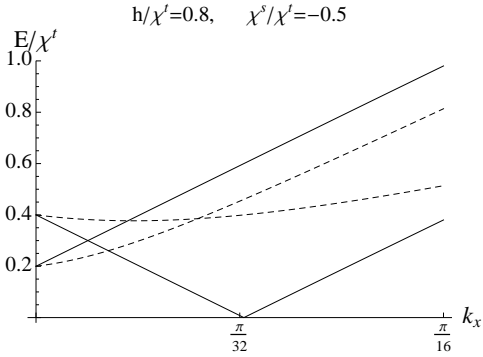}
\caption{Detail of the single-particle excitation spectrum of the MB1-Dirac state for $\chi^s / \chi^t = -0.5$ and $h / \chi^t = 0.8$.  The plot is again along the line $k_y = 0$, near the node at $k_x = 0$.  The solid lines are the energies in the $t^1$-$t^2$ sector, and the dashed lines the energies in the $s$-$t^3$ sector.  It is evident that a gap has opened in the $s$-$t^3$ sector.  The spectrum away from the nodes is very similar to the case $\chi_s / \chi_t = 0.5$, as plotted in Fig.~\ref{fig:cspos}.}
\label{fig:csneg_small}
\end{figure}

We now return to the lattice to discuss the effect of Zeeman magnetic field.  In the low-energy effective theory, a $z$-axis Zeeman field gives the following contribution to the Hamiltonian:
\begin{equation}
H_{{\rm Zeeman}} = h \frac{i c_{st}}{4} \sum_{\br} s_{\br} t^3_{\br} +  h \frac{i c_{tt}}{4} \sum_{\br} t^1_{\br} t^2_{\br} \text{,}
\end{equation}
where $c_{st}$ and $c_{tt}$ are dimensionless, non-universal constants.  Two terms appear in this Hamiltonian because each of these terms has precisely the same symmetries as the microscopic magnetic field operator $\sum_{\br} S^z_{\br}$.  In the presence of one of these terms (say $c_{st} = 0$ but $c_{tt} \neq 0$) and short-range fermion interactions, we expect the other term will be generated. Therefore, both terms must be included.

We focus first on the sector of the theory involving $s$- and $t^3$-fermions.  Neglecting coupling to the gauge field and fermion interactions (which, for the effect of a small Zeeman field on the low-energy spectrum, is not an approximation), the single-particle excitation spectrum is given by
\begin{equation}
\epsilon^{st}_{\pm}(\bk) = \Big| (\chi^s + \chi^t) f_{\bk} \pm \sqrt{ (\chi^s - \chi^t)^2 f_{\bk}^2 + \tilde{h}_{st}^2 } \Big| \text{,} \label{eqn:st-spectrum}
\end{equation}
where we have defined $\tilde{h}_{st} = h c_{st} / 4$ and
\begin{equation}
f_{\bk} =  \sqrt{4 - 2 \cos(2 k_x) - 2 \cos(2 k_y) } \text{.}
\end{equation}
Each energy level is two-fold degenerate, and it should be recalled that  that $\bk$ and $-\bk$ are equivalent points.  The spectrum in the sector involving $t^1$- and $t^2$-fermions can be obtained from Eq.~(\ref{eqn:st-spectrum}) by putting $\chi^s \to \chi^t$ and $\tilde{h}_{st} \to \tilde{h}_{tt} = h c_{tt} / 4$.

The effect of the Zeeman field differs considerably depending on whether $\chi_s$ and $\chi_t$ have the same or opposite signs.  When $\chi_s / \chi_t > 0$, a small Zeeman field opens  Fermi pockets around the Dirac nodes in both the $s$-$t^3$ and $t^1$-$t^2$ sectors.  On the other hand, when $\chi_s / \chi_t < 0$, a small Zeeman field opens a gap in the $s$-$t^3$ sector.  This is illustrated in Figures~\ref{fig:zerob}-\ref{fig:csneg_small} (in all cases shown we choose $c_{st} = c_{tt} = 1$).  These spectral features will be manifest in the spin structure factor $S(\bq,\omega)$, which can be measured by neutron scattering.  Calculation of $S(\bq,\omega)$ for the MB1-Dirac state is left for future work.  We note that similar interesting behavior of $S = 1$ Majorana spinons in a Zeeman field was also found in the exactly solvable model of Ref.~\onlinecite{lai11}.

Finally, returning to the continuum effective theory, we consider the properties of some gapped phases nearby to the MB1-Dirac state.  In particular, given the appearance of non-Abelian statistics in the $B$-phase of the Kitaev honeycomb lattice model,\cite{kitaev06} as well as in one of the exactly solvable models with $S = 1$ Majorana spinons,\cite{yao11} it is natural to ask whether such statistics can also occur upon opening a gap in the MB1-Dirac state.  We answer this question in the affirmative below, although -- at least if we maintain spin rotation symmetry -- it seems to be necessary \emph{both} to induce a dimerization of the square lattice, \emph{and} to break time-reversal symmetry.  It is likely that there are other routes to non-Abelian statistics if we allow for breaking of spin rotation symmetry, but we have not studied this possibility.

In the gapped phases we consider, various symmetries are broken -- this symmetry breaking can be induced explicitly, or could potentially be induced spontaneously as a result of sufficiently large fermion interactions.  In all these phases, $Z_2$ vortex excitations remain gapped and the $Z_2$ gauge field remains in its deconfined phase.  In particular, we consider the following perburbation to the Lagrangian:
\begin{equation}
\delta {\cal L} = i m^s_{C} \bar{\Psi} \Psi + i m^t_C \bar{\Phi}^i \Phi^i + i m^s_D \bar{\Psi} \mu^3 \Psi +  i m^t_D \bar{\Phi}^i \mu^3 \Phi^i \text{.}
\end{equation}
For simplicity of discussion, we restrict attention to the case $m^s_C, m^t_C, m^s_D, m^t_D > 0$.  As discussed in Ref.~\onlinecite{kitaev06}, the $Z_2$ vortices in a system such as this one have non-Abelian statistics when the Chern number $\nu$ is odd, stemming from the binding of an odd number of Majorana modes to each vortex.

First we consider the case $m^s_C, m^t_C \neq 0$ and $m^s_D, m^t_D = 0$, where the Lagrangian breaks parity and time reversal symmetries, while preserving the product of parity and time reversal, as well as other microscopic symmetries.  In this case each species of  fermion on the lattice has Chern number $\nu = 1$, for a total Chern number $\nu = 4$, giving rise to four gapless, co-propagating Majorana edge modes.  (We note that if we allow $m^s_C$ and $m^t_C$ to have opposite sign, we can also obtain $\nu = 2$.)  Because the Chern number is even, the vortices do not have non-Abelian statistics.

Next, we consider $m^s_C, m^t_C = 0$ and $m^s_D, m^t_D \neq 0$.  This corresponds to inducing a columnar dimerization of the square lattice.  Each lattice fermion has $\nu = 0$, so the total Chern number $\nu = 0$, and the fermion spectrum in this state is topologically trivial.  In particular there are neither gapless edge states nor non-Abelian statistics.

Finally, we consider the case where all the mass terms are nonzero, which corresponds to co-existing columnar dimer and time-reversal-breaking orders.  First we consider the $s$-fermion sector.  As long as $ m^s_C  > m^s_D$, then $\nu = 1$.  When $ m^s_C  = m^s_D$, the gap closes, and for $ m^s_C  < m^s_D$ we obtain a topologically trivial state with $\nu = 0$.  The same statements hold for the $t$-fermion sector, except that the total $t$-fermion Chern number when $m^t_C > m^t_D$ is $\nu = 3$.  Based on this, we note that if $m^s_C  >  m^s_D$ but $m^t_C  < m^t_D$, we have a total Chern number $\nu = 1$, giving rise to one gapless chiral Majorana edge mode.  If we reverse the above inequalities, we instead have $\nu = 3$, with three Majorana edge modes.  Non-Abelian statistics arise in both these cases.

\section{Discussion}
\label{sec:discussion}

In this paper, we showed that, in the fermionic parton approach to $S = 1/2$ spin liquids with ${\rm SU}(2)$ symmetry, spin rotations can be realized in two distinct ways.  In addition to the familiar naive realization, where spin rotations act on the $S = 1/2$ partons with no accompanying gauge transformation, we also found a projective realization, where ${\rm SU}(2)$ spin and gauge rotations are locked together.  This projective realization leads to spin liquids with $S = 1$ and $S = 0$ Majorana fermion excitations coupled to a deconfined $Z_2$ gauge field.  We discussed the projective symmetry group (PSG) classification of such states, showing that their PSGs are in one-to-one correspondence with ${\rm SU}(2)$ PSGs in the existing classification.  To illustrate these results we studied states in each of the four Majorana PSGs on the two-dimensional square lattice, finding that one, the MB1-Dirac state, is a stable phase with gapless Fermi points.  This phase exhibits interesting behavior in a Zeeman magnetic field, and has nearby gapped phases supporting non-Abelian statistics.

We now discuss the challenging and very interesting question of where  Majorana spin liquids may be found in models (beyond the exactly solvable models of Refs.~\onlinecite{wangf10, yao11, lai11, lai11b}) and in real systems.  In Sec.~\ref{sec:frustrated}, we noted that frustrated mean-field ans\"{a}tze for Majorana spin liquids break time reversal symmetry.  It is likely that such time-reversal-breaking states will have lower energies in frustrated $S = 1/2$ magnets than Majorana spin liquids with time reversal symmetry, because they may be better able to gain exchange energy from all the bonds of the frustrated lattice.  Because frustration is expected to be an important ingredient in stabilizing spin liquid phases, it may be more productive to search for time-reversal-breaking Majorana spin liquids than their fully symmetric cousins.  It will be interesting to classify and study such states in future work.

Another promising place to look for spin liquids in general is in weak Mott insulators; that is, just on the insulating side of a Mott metal-insulator transition.\cite{balents10}  A given spin liquid state is naturally associated with a corresponding state, not necessarily unique, obtained by condensing charge-carrying excitations in the spin liquid.  This state can be a metal, superconductor or band insulator.  This may also be a promising place to look for Majorana spin liquids;  indeed, a Majorana spin liquid with projective spin rotation symmetry has already been discussed in a continuum effective theory of the honeycomb lattice Hubbard model (phase B2 of Ref.~\onlinecite{xu10}).  

Majorana spin liquids could be studied in Hubbard models using the ${\rm SU}(2)$ slave-rotor approach,\cite{hermele07, kskim06, kskim07} which is an extension to  the Hubbard model of the ${\rm SU}(2)$ gauge theory of the Heisenberg model.  For single band Hubbard models at half-filling on bipartite lattices, in addition to the ${\rm SU}(2)$ spin rotation symmetry, there is a ${\rm SU}(2)$ pseudospin symmetry, of which ${\rm U}(1)$ charge rotations is a subgroup.\cite{cnyang89,zhang90}  In the slave rotor framework, we expect that the state obtained from a Majorana spin liquid by condensing charge-carrying excitations will have a spontaneous locking of spin and pseudospin symmetries; that is, ${\rm SU}(2) \times {\rm SU}(2)$ symmetry is broken to the diagonal ${\rm SU}(2)$ subgroup of simultaneous spin and pseudospin rotations.  This expectation is corroborated by the results of Ref.~\onlinecite{xu10}. (There, phase B2 is adjacent to phase B, which has such a ${\rm SU}(2) \times {\rm SU}(2) \to {\rm SU}(2)$ symmetry breaking.)  In more realistic situations, the ${\rm SU}(2)$ pseudospin is broken down to ${\rm U}(1)$ charge rotations, and we may expect instead a locking of charge rotations to a ${\rm U}(1)$ subgroup of spin rotations, corresponding to breaking ${\rm SU}(2) \times {\rm U}(1) \to {\rm U}(1)$.

On a different note, it is interesting to ask whether our starting point of $S = 1/2$ fermionic partons is necessary to describe the Majorana spin liquids discussed here -- it is not.  We could have introduced Majorana partons from the beginning, using the representation of the spin operator in Eq.~(\ref{eqn:s-in-majoranas}) and the constraint Eq.~(\ref{eqn:majorana-constraint}).  Considering mean-field Hamiltonians of the form Eq.~(\ref{eqn:majorana-h0}), we would again conclude that $T^i$ [as defined in Eq.~(\ref{eqn:t-in-majoranas})] and not $S^i$ commutes with the Hamiltonian.  The status of spin rotations in such a state is potentially a confusing issue, but the results of this paper -- and the starting point of $S = 1/2$ partons -- clarify that this state is in fact spin-rotation invariant.

Actually there is some freedom in the expression of $S^i_{\br}$ in terms of Majorana partons; more generally, we may have
\begin{equation}
S^i_{\br} = - \frac{1}{4} \big[ (1-x) i s_{\br} t_{\br i} + (1+x) \frac{i}{2} \epsilon_{i j k} t_{\br j} t_{\br k} \big] \text{.}
\end{equation}
Setting $x=0$ gives the representation we obtained from $S = 1/2$ fermions, and $x = -1$ is the representation used by Kitaev to solve his honeycomb lattice model.\cite{kitaev06}  Putting $x=1$ gives $S^i_{\br} = T^i_{\br}$, as in the representation developed in Refs.~\onlinecite{tsvelik92,coleman93,shastry97}, and employed by Biswas \emph{et. al.}\cite{biswas11} (see Appendix~\ref{app:bfls}).  Unlike in that representation, however, even when $x=1$ there is still an $s$-fermion that enters via the constraint.

We close with a discussion of some open issues for future study.  It would be interesting if the formalism developed here can reproduce the $S = 1$ Majorana spin liquid phases of the exactly solvable models of Refs.~\onlinecite{wangf10, yao11, lai11, lai11b}.  We have made some attempts in this direction for the model of Ref.~\onlinecite{yao11}, but so far have not been successful.  It would also be interesting to study projected wavefunctions for Majorana spin liquids, which we discussed briefly in Sec.~\ref{sec:majorana}.  Finally, there are no doubt other circumstances where a continuous global symmetry is realized projectively in a spin liquid or other exotic phase.  For example, suppose we consider a $S = 1/2$ system with only global ${\rm U}(1)$ symmetry, which we can think of as a system of strongly correlated bosons.  Treating this system using $S = 1/2$ fermionic partons, the ${\rm U}(1)$ symmetry can be realized projectively by a locking to a ${\rm U}(1)$ subgroup of the ${\rm SU}(2)$ gauge group.  Due to the lower symmetry, we expect that the form of the mean-field Hamiltonian is less constrained, and that a wider variety of spin liquid states may occur.   Moreover, projective continuous symmetry can be realized in this system with \emph{bosonic} $S = 1/2$ partons, where the gauge group is ${\rm U}(1)$, by a locking of gauge and global ${\rm U}(1)$ symmetries.  These and other similar states may have unexpected properties, and may prove important in broadening the understanding of exotic states of matter.

While this paper was being finalized, we learned that T. Senthil has independently obtained some of the main results presented here.\cite{senthilunpub}

\begin{acknowledgments}
We are grateful to Subir Sachdev for useful discussions.  This research is supported by the David and Lucile Packard Foundation.
\end{acknowledgments}

\appendix

\section{Restrictions on $M_{\br}$}
\label{app:mmatrix}

In Sec.~\ref{sec:projective}, it is asserted that, starting from the form $T^i_{\br} = S^i_{\br} + M_{\br}^{i j} G^j_{\br}$ (with $M^{i j}_{\br}$ an arbitrary real $3 \times 3$ matrix), the requirement $[ T^i_{\br} , T^j_{\br} ] = i \epsilon^{i j k} T^k_{\br}$ implies either $M_{\br} = 0$ or $M_{\br} \in {\rm SO}(3)$.  In this Appendix we prove this assertion.  

The required commutation relation implies the following equation for the matrix $M$ (here and below, we drop the site label $\br$):
\begin{equation}
\epsilon^{i j k} M^{k m} = M^{i l} M^{j k} \epsilon^{l k m} \text{.}
\label{eqn:m-equation}
\end{equation}
We will now find all possible solutions of this equation for $M$.  We proceed by making the singular value decomposition $M = \sigma U D V$, where $\sigma = \pm 1$, $U, V \in {\rm SO}(3)$, and $D = \operatorname{diag}(d_1, d_2, d_3)$, where $d_i \geq 0$.  Making use of the identities $U^{i j} U^{k j} = \delta^{i k}$ and $U^{i i'} U^{j j'} U^{k k'} \epsilon^{i' j' k'} = \epsilon^{i j k}$, and similarly for $V$, Eq.~(\ref{eqn:m-equation}) can be brought to the form
\begin{equation}
\sigma \epsilon^{i j m} D^{m k} = D^{i n} D^{j m} \epsilon^{n m k} \text{.}
\end{equation}
This gives the equation $\sigma d_3 = d_1 d_2$ and its two cyclic permutations.  Clearly one solution is $d_i = 0$, corresponding to $M = 0$.  If any one of the $d_i$ is nonzero, then clearly they must all be nonzero.  Moreover, in the case of a non-zero solution, since the $d_i$ are positive, we must have $\sigma = 1$.  It is then trivial to show that the only non-zero solution is $d_1 = d_2 = d_3 = 1$, which corresponds to $M \in {\rm SO}(3)$.

\section{Invariant gauge group and form of symmetry operations}
\label{app:igg}

Here, we show that Majorana spin liquid mean-field ans\"{a}tze, with $H_0$ as given in Eq.~(\ref{eqn:majorana-h0-Fs}), have $\text{IGG} = Z_2$.  This statement holds as long as  $\chi^2_{\br \br'} \neq 0$ on some bonds -- we recall that we always assume this to be the case, since if $\chi^2_{\br \br'} = 0$ everywhere there is no distinction between naive and projective spin rotations, and we simply have an ansatz with $\text{IGG} = {\rm SU}(2)$.  We also show that the space group and time reversal operations leaving $H_0$ invariant take a simple form.

To show that $\text{IGG} = Z_2$,  consider first any two sites $\br$ 
and $\br'$ for which $\chi^2_{\br \br'}$ is nonzero.  The Hamiltonian for this bond is
\begin{equation}
H_{\br \br'} =  i \chi^1_{\br \br'} \operatorname{tr} (F^{\vphantom\dagger}_{\br} F^\dagger_{\br'} )
+ i \chi^2_{\br \br'} \operatorname{tr} ( \sigma^i F^{\vphantom\dagger}_{\br} \sigma^i F^\dagger_{\br'} ) \text{.}
\end{equation}
We wish to find all gauge transformations $F_{\br} \to F_{\br} U_{\br}$ leaving $H_{\br \br'}$ invariant.  It is useful at this point to note that the 16 fermion bilinears $\{ \operatorname{tr} (F_{\br} F^\dagger_{\br'} ) ,
\operatorname{tr}  ( \sigma^i F_{\br} F^\dagger_{\br'} ) ,
\operatorname{tr}  (F_{\br}  \sigma^i F^\dagger_{\br'} ) ,
\operatorname{tr}  ( \sigma^i F_{\br} \sigma^j F^\dagger_{\br'} ) \}$ form a complete, linearly independent set of all fermion bilinears connecting the two sites $\br$ and $\br'$.  In particular, this means that any gauge transformation leaving $H_{\br \br'}$ invariant must separately leave the $\chi^1$ and $\chi^2$ terms invariant: gauge transformations send the $\chi^1$ term to a linear combination of the bilinears $\{  \operatorname{tr} (F_{\br} F^\dagger_{\br'} ) , \operatorname{tr}  (F_{\br}  \sigma^i F^\dagger_{\br'} ) \}$, while the $\chi^2$ term becomes a linear combination in the distinct subspace of bilinears spanned by $\{ \operatorname{tr}  ( \sigma^i F_{\br} F^\dagger_{\br'} ) , \operatorname{tr}  ( \sigma^i F_{\br} \sigma^j F^\dagger_{\br'} ) \}$.

In order for the $\chi^2$ term to be invariant, we must have
\begin{equation}
U_{\br} \sigma^i U^\dagger_{\br'} = \sigma^i \text{,}
\end{equation}
for all $i = 1,2,3$.  The only solution to this equation is $U_{\br} = U^\dagger_{\br'} = \pm 1$.  The $\chi^1$ term is also invariant under such a gauge transformation, so we do not need to consider it separately.

Next we need to go beyond the original two sites $\br$ and $\br'$.  Suppose there is a third site $\br''$ connected to $\br$ such that  $\chi^1_{\br \br''} \neq 0$ or $\chi^2_{\br \br''} \neq 0$.  In order to leave $H_{\br \br''}$ invariant, It is easy to see that we must have $U_{\br''} = U_{\br}$.  Because we assume the ansatz to be connected, we can repeat this procedure to determine $U_{\br}$ for the entire lattice.  Therefore we have shown that the only gauge transformations leaving $H_0$ invariant are $U_{\br} = 1$ for all $\br$, and $U_{\br} = -1$ for all $\br$, and thus $\text{IGG} = Z_2$.

Now we discuss some restrictions on the form of symmetry operations.  As usual, we suppose that $H_0$ is invariant under a set of space group symmetries labeled by $S$, and also under time reversal symmetry ${\cal T}$.  As we see below, these operations must take the form
\begin{eqnarray}
S : F_{\br} &\to& F_{S(\br)} \pi^S_{\br} \label{eqn:app-general-sg-transformation} \\
{\cal T} : F_{\br} &\to& (i \sigma^2) F_{\br} (i \sigma^2) \pi^{\cal T}_{\br}  \text{,} \label{eqn:app-general-t-transformation}
\end{eqnarray}
where $\pi^S_{\br}$ and $\pi^{\cal T}_{\br}$ take values $\pm 1$ as a function of lattice site $\br$.

We first consider the space group transformations.  The most general action of a space group transformation $S$ on fermion operators is
\begin{equation}
S : F_{\br} \to F_{S(\br)} U^S_{\br} \text{.}
\end{equation}
We can think of this operation as a composition of the gauge transformation $U^S_{\br}$ followed by the operation $F_{\br} \to F_{S(\br)}$.  It is useful to note that the gauge transformation $U^S_{\br}$ need not leave $H_0$ invariant, but must transform the Hamiltonian on each bond $H_{\br \br'}$ into another Hamiltonian \emph{of the same form}, meaning that the result of the gauge transformation can be absorbed into a change of $\chi^1_{\br \br'}$ and $\chi^2_{\br \br'}$.  For two sites $\br$ and $\br'$ with $\chi^2_{\br \br'} \neq 0$, the only such transformations are $U^S_{\br} = U^S_{\br'} = \pm 1$, and $U^S_{\br} = - U^S_{\br'} = \pm 1$.  Now, as above in determining the IGG, suppose a third site $\br''$ is connected to $\br$ such that  $\chi^1_{\br \br''} \neq 0$ or $\chi^2_{\br \br''} \neq 0$.  In order for $H_{\br \br''}$ to transform into another bond Hamiltonian of the same form, we must have $U^S_{\br''} = \pm 1$.  Following this procedure to extend the transformation to the whole lattice, we find $U_{\br}$ can only take values $\pm 1$ for all $\br$, and thus Eq.~(\ref{eqn:app-general-sg-transformation}) holds.

Next, we consider time reversal.  The most general realization of time reversal is
\begin{equation}
{\cal T} : F_{\br} \to (i \sigma^2) F_{\br} U^{\cal T}_{\br} (i \sigma^2) \text{,}
\end{equation}
where the presence of $(i \sigma^2)$ on the right is a convention, and it should be kept in mind that the transformation is anti-unitary.  This transformation must leave $H_{\br \br'}$ invariant.  Acting on $H_{\br \br'}$, we have
\begin{eqnarray}
{\cal T} : H_{\br \br'} &\to& -i \chi^1_{\br \br'} \operatorname{tr} (F^{\vphantom\dagger}_{\br} U^{\cal T}_{\br} ( U^{\cal T}_{\br'} )^\dagger F^\dagger_{\br'} ) \nonumber \\
&-& i \chi^2_{\br \br'} \operatorname{tr} ( \sigma^i F_{\br} U^{\cal T}_{\br} \sigma^i (U^{\cal T}_{\br'} )^\dagger F^\dagger_{\br'} )
\end{eqnarray}
As long as $\chi^2_{\br \br'} \neq 0$, the only transformations leaving $H_{\br \br'}$ invariant are $U^{\cal T}_{\br} = - U^{\cal T}_{\br'} = \pm 1$.  We can follow the above procedure to extend $U^{\cal T}_{\br}$ to the whole lattice, and find that $U^{\cal T}_{\br}$ can only take values $\pm 1$ for all $\br$, thus showing Eq.~(\ref{eqn:app-general-t-transformation}).

\section{Continuum field theory for MB1-Dirac state}
\label{app:mb1}

Here, we work out the low-energy effective theory for the MB1-Dirac state, at the mean-field level, and quote the action of the microscopic symmetry operations in the low-energy fermion fields.  As discussed in Sec.~\ref{sec:mb1-properties}, this low-energy theory is also valid beyond mean-field theory.

We begin by working with the  $s$-fermion part of the mean-field Hamiltonian.  Because the development for the $t$-fermions exactly parallels the treatment below, there is no need to go through it explicitly.
We have
\begin{equation}
H_{0s} = \chi^s \sum_{\bk} m_{i j}(\bk) s_{\bk i} s_{-\bk j} \text{,}
\end{equation}
where the sum is over the Brillouin zone $| k_x | \leq \pi$, $| k_y | \leq \pi/2$,
and
\begin{equation}
m(\bk) 
= \left( \begin{array}{cc}
2 \sin(k_x) & i ( 1 - e^{2 i k_y} ) \\
-i (1 - e^{-2 i k_y}) & - 2 \sin(k_x)
\end{array} \right)  \text{.}
\end{equation}
The nodes are at $\bk = 0$ and $\bk = \bK = (\pi, 0)$.  Letting $\bq$ be small on the scale of the zone size, we have
\begin{eqnarray}
m(\bq) &=& 2 q_x \tau^3 + 2 q_y \tau^1 \\
m(\bq + \bK) &=& - 2 q_x \tau^3 + 2 q_y \tau^1 \text{.}
\end{eqnarray}

We define continuum fields by writing
\begin{eqnarray}
\tilde{\psi}_1 &\sim& \left( \begin{array}{cc} s_{\bq 1} \\ s_{\bq 2} \end{array} \right) \\
\tilde{\psi}_2 &\sim& \tau^1 \left( \begin{array}{cc} s_{\bq + \bK , 1} \\ s_{\bq + \bK, 2} \end{array} \right) \text{,}
\end{eqnarray}
where $\tau^i$ are $2 \times 2$ Pauli matrices.  These fields obey the continuum Hamiltonian density
\begin{eqnarray}
{\cal H}_{0s} &=&  v_s \psi^T_1 ( i \tau^3 \partial_x + i \tau^1 \partial_y) \psi_1 \nonumber \\
&+&   v_s \psi^T_2 ( i \tau^3 \partial_x + i \tau^1 \partial_y) \psi_2 \text{,}
\end{eqnarray}
where we have introduced the nodal velocity $v_s$.
Defining $\bar{\psi}_1 = \psi^T_1 (- i \tau^2)$ and $\bar{\psi}_2 = \psi^T_2 (- i \tau^2)$, the imaginary-time Lagrangian density is
\begin{equation}
{\cal L}_s = \bar{\psi}_1[  i \gamma_{\mu} \partial^s_{\mu} ] \psi_1 +  \bar{\psi}_2 [ i \gamma_{\mu} \partial^s_{\mu} ]  \psi_2 \text{,}
\end{equation}
where $\mu = 0,1,2$,
\begin{equation}
\partial^s_{\mu} = (\partial_0, v_s \partial_1 , v_s \partial_2 ) \text{,}
\end{equation}
and
\begin{equation}
\gamma_{\mu} = (\tau^2, \tau^1, -\tau^3) \text{.}
\end{equation}
We define the four-component spinor
\begin{equation}
\Psi = \left(\begin{array}{c} \psi_1 \\ \psi_2 \end{array}\right) \text{,}
\end{equation}
and define $\mu^i$ Pauli matrices acting in the flavor space mixing $\psi_1$ and $\psi_2$.  For example, 
\begin{equation}
\mu^3 \Psi = \left( \begin{array}{c} \psi_1 \\  -\psi_2 \end{array}\right) \text{.}
\end{equation}

For each triplet fermion $t^i$, we proceed identically and introduce a  four-component spinor field $\Phi^i$.  The full continuum Lagrangian density is then 
\begin{equation}
{\cal L} = \bar{\Psi} [ i \gamma_{\mu} \partial^s_{\mu} ] \Psi + 
\bar{\Phi}^i [ i \gamma_{\mu} \partial^t_{\mu} ] \Phi^i \text{,}
\end{equation}
where
\begin{equation}
\partial^t_{\mu} = (\partial_0, v_t \partial_1, v_t \partial_2 ) \text{,}
\end{equation} 
and we have introduced the nodal velocity $v_t$ of the $t$-fermions.

The action of space group and time reversal symmetries on the lattice fermion fields is given in Eqs.~(\ref{eqn:mb1-psg-first}-\ref{eqn:mb1-psg-last}).  It is straightforward but somewhat tedious to work out the action of these operations on the continuum fields.  As these symmetries act identically on $\Psi$ and $\Phi^i$, we now quote their action on $\Psi$:
\begin{eqnarray}
T_x : \Psi(\br) &\to& \mu^3 \Psi(\br) \\
T_y : \Psi(\br) &\to& \mu^1 \Psi(\br) \\
P_x : \Psi(\br) &\to& \tau^1 \mu^1 \Psi(\br') \\
P_{xy} : \Psi(\br) &\to& \frac{1}{2} (\tau^1 + \tau^3) (\mu^1 + \mu^3) \Psi(\br') \\
{\cal T} : \Psi(\br) &\to& (i \tau^2) (i \mu^2) \Psi(\br) \text{.}
\end{eqnarray}
Here, $\br = (r_x, r_y)$, $P_x : \br \to \br' = (-r_x,r_y)$, and $P_{xy} : \br \to \br' = (r_y, r_x)$.

\section{Approach of Biswas \emph{et. al.}}
\label{app:bfls}

A different approach to $S = 1$ Majorana spin liquids was introduced by Biswas, Fu, Laumann and Sachdev (BFLS),\cite{biswas11} where a mean-field theory of a particular Majorana spin liquid on the triangular lattice (BFLS state) was constructed.  This approach has some similarities to ours, but differs in the absence of a singlet $s$-fermion.  Because we feel it may be useful for future work on Majorana spin liquids, we elaborate here on the differences between our approach and that of BFLS.  In particular, we discuss how fluctuations beyond mean-field theory may be incorporated in the approach of BFLS.  We argue that following the standard route to construct an effective lattice gauge theory -- a route that is successful in our formalism -- is problematic, except when symmetry allows for a partition of the lattice into dimers.  Similar issues arise in constructing a projected wavefunction.  However, we discuss a different route by which fluctuations may be included,\cite{shastry97, sachdevpc} which appears to be sound.  This approach has, to our knowledge, not been studied in detail, and it will be very interesting to do so in future work.  The question of constructing a projected wavefunction, starting from the mean-field theory of the BFLS state, is still open.

These differences notwithstanding, our approach can describe a very similar state to that of BFLS, differing only in the presence of a gapless $s$-fermion at the mean-field level.  This state and the BFLS state are expected to have very similar physical properties.  Indeed, it is conceivable that these two mean-field theories are different limits of the same phase.

\begin{figure}
\includegraphics[width=1.5in]{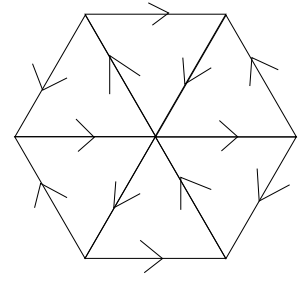}
\caption{Orientation of nearest-neighbor bonds on the triangular lattice, used to define the mean-field Hamiltonians of Eqs.~(\ref{eqn:h0-bfls}) and~(\ref{eqn:h0-bfls2}).  This pattern of orientations respects the full translation symmetry of the triangular lattice.}
\label{fig:bfls}
\end{figure}

The starting point of Ref.~\onlinecite{biswas11} is a parton representation developed in Refs.~\onlinecite{tsvelik92,coleman93,shastry97}, where the spin operators on each lattice site are represented directly using a triplet of $S = 1$ Majorana fermions:
\begin{equation}
S^i_{\br} = - \frac{i}{4} \epsilon^{i j k} t^j_{\br} t^k_{\br} \text{,}
\label{eqn:shastry-sen}
\end{equation}
where $\{ t^i_{\br} , t^j_{\br'} \} = 2 \delta^{i j} \delta_{\br \br'}$.  Focusing here and below on the triangular lattice and the BFLS state, these fermions are taken to obey a mean-field Hamiltonian
\begin{equation}
H_0 = i \chi^t \sum_{\langle \br \br' \rangle}   \bt_{\br} \cdot \bt_{\br'} \text{,}
\label{eqn:h0-bfls}
\end{equation}
where the orientations of nearest-neighbor bonds $\langle \br \br' \rangle$ are chosen as shown in Fig.~\ref{fig:bfls}.  As noted in Ref.~\onlinecite{biswas11}, this mean-field Hamiltonian preserves the full translation symmetry of the triangular lattice, but breaks time-reversal and certain point group symmetries.  The fermions have an interesting locus of gapless excitations in momentum space.\cite{biswas11}  In particular, the fermions are gapless at the $\Gamma$ point of the Brillouin zone, and the vanishing gap at this point is protected by translation symmetry.

As usual, Eq.~\ref{eqn:shastry-sen} itself does not completely define the parton representation of the spin model; the physical Hilbert space must also be specified as a constraint on the larger parton Hilbert space.  Because there are an odd number of Majorana fermions on each lattice site, it is not possible to define a parton Hilbert space for individual lattice sites.  Ref.~\onlinecite{biswas11} describes how one may proceed:  the lattice can be (arbitrarily) partitioned into dimers.  Suppose that $\br_1$ and $\br_2$ are the two sites in some particular dimer.  We can define three complex fermion operators by writing
\begin{equation}
c^i_{\br_1 \br_2} = \frac{1}{2} ( t^i_{\br_1} + i t^i_{\br_2} ) \text{,}
\end{equation}
which satisfy anticommutation relations $\{ c^i_{\br_1 \br_2} , ( c^j_{\br_1 \br_2} )^\dagger \} = \delta^{i j}$.  We can choose the physical Hilbert space of the dimer to be those states with an even number of $c^i$-fermions; the state with no fermions is the spin singlet, and the three states with two fermions form the spin triplet.  This choice is enforced by the local $Z_2$ constraint
\begin{equation}
D_{\br_1 \br_2} \equiv - i t^1_{\br_1} t^2_{\br_1} t^3_{\br_1} t^1_{\br_2} t^2_{\br_2} t^3_{\br_2} = 1 \text{.} 
\end{equation}
It should be noted that the choice $D_{\br_1 \br_2} = -1$ is also legitimate; this picks out the states with an odd number of $c^i$-fermions, where the single-fermion states make up a spin triplet, and the unique three-fermion state is a spin singlet.  These two choices of constraint are in fact interchanged by the $Z_2$ ``gauge transformation'' $t^i_{\br_1} \to t^i_{\br_1}$, $t^i_{\br_2} \to - t^i_{\br_2}$.  This transformation does not satisfy the usual requirement for a gauge transformation that it leave the physical Hilbert space invariant.  There is also a more conventional $Z_2$ gauge freedom under transformations $t^i_{\br} \to \pi_{\br} t^i_{\br}$, where $\pi_{\br}$ takes values $\pm 1$ and is constant on each dimer.

For a given partition of the lattice into dimers, we can specify a sector of Hilbert space by choosing $D_{\br_1 \br_2} = \pm 1$ on each dimer.  Importantly, because the spin operators are invariant under the $Z_2$ transformations that change the constraint, the physics is the same in every sector.  That is, all correlation functions of spin operators are the same in every sector.  

So far, we have described an exact parton representation of the spin model, with no approximations.  Now, we consider the mean-field starting point of $H_0$, and discuss how fluctuations may be included.  We shall first follow the standard route to construct an effective lattice gauge theory, and show that this approach is problematic, except when symmetry allows for a natural partition of the lattice into dimers.  

We begin with a particular sector of the parton Hilbert space, with a fixed set of local constraints.  For simplicity of discussion we choose $D_{\br_1 \br_2} = 1$ on each dimer.  Those $Z_2$ transformations preserving the constraint (\emph{i.e.}, those transformations that are constant on dimers) play the role of gauge transformations.  With this in mind, we should define $Z_2$ gauge fields $\sigma^z$ and $\sigma^x$, not on the links of the original lattice, but on links connecting nearby dimers.  The local constraint $D_{\br_1 \br_2} = 1$ is promoted to the gauge constraint $D_{\br_1 \br_2} = \prod \sigma^x$, where the product is over those $Z_2$ electric fields $\sigma^x$ touching the dimer $(\br_1, \br_2)$.  Hopping terms within a dimer are unchanged from $H_0$, but those between dimers are multiplied by the $Z_2$ vector potential $\sigma^z$.  Because any dimer covering of the triangular lattice breaks translation symmetry, it is clear that this effective gauge theory has less symmetry than $H_0$, and moreover the breaking of lattice symmetry depends on the dimer covering chosen.  Therefore the mean-field Hamiltonian $H_0$ does not properly capture the physics described by the more complete effective lattice gauge theory.  Of course, if $H_0$ itself breaks translation symmetry in such a way that there is a natural dimer covering, then we can employ this dimer covering in the construction of the effective gauge theory, and no problems arise.  

The issue that arose here is essentially that, while the parton representation employed in Ref.~\onlinecite{biswas11} gives an exact representation of the spin model when the constraint $D_{\br_1 \br_2} = 1$ is imposed exactly, some symmetries are explicitly broken the moment the constraint is softened, as occurs in the route we described to construct an effective lattice gauge theory.  A very similar issue arises and has been discussed in ${\rm U}(1)$ slave-rotor theories of the Hubbard model on bipartite lattices (see Sec. IV.A of Ref.~\onlinecite{hermele07}).

Similar issues arise in the construction of a projected wavefunction from the mean-field starting point of $H_0$.  Given a partition of the lattice into dimers labeled by $D$, we define ${\cal P}_D$ to project onto the physical Hilbert space of each dimer.  Then a projected wavefunction is given by $| \psi \rangle = {\cal P} | \psi_0 \rangle$, where $| \psi_0 \rangle$ is the ground state of $H_0$, and ${\cal P} = \prod_D {\cal P}_D$.  Because ${\cal P}$ does not commute with translations, we expect that $|\psi\rangle$ is not invariant under translations and thus has less symmetry than $H_0$.  Once again, if translations are broken in $H_0$ so that there is a preferred dimer covering, this dimer covering can be used to construct ${\cal P}$, and no issue arises.

Despite these difficulties, fluctuations can be included by a different route that does not suffer from the above issues.\cite{shastry97,sachdevpc}  We can begin with the Heisenberg spin model, represented in terms of partons using Eq.~(\ref{eqn:shastry-sen}), but without any local constraints.  In the resulting Grassmann functional integral, we have only the time-derivative term for the fermions, and the quartic spin-spin interaction.  To obtain this description we have summed over all the sectors of Hilbert space described above; this is legitimate because the physics is the same in every sector.  The spin-spin interaction can then be decoupled by standard means and $H_0$ can be obtained as a mean-field saddle point.  To our knowledge, fluctuations about this mean-field saddle point, or, indeed, about any saddle point in this construction, have not yet been studied.  It will be important to investigate this in future work.  Moreover, it may be possible to use these ideas to construct a projected wavefunction for the BFLS state that does not suffer from the issues described above.

Finally, we turn to the description of a state very similar to the BFLS state, using our formalism.  We consider the mean-field Hamiltonian 
\begin{equation}
H_0 = i \chi^t \sum_{\langle \br \br' \rangle}  \bt_{\br} \cdot \bt_{\br'} +  i \chi^s \sum_{\langle \br \br' \rangle}   s_{\br}  s_{\br'}  \text{,}
\label{eqn:h0-bfls2}
\end{equation}
where again the orientations of nearest-neighbor bonds $\langle \br \br' \rangle$ are chosen as in Fig.~\ref{fig:bfls}.  If it were possible to gap out the $s$-fermions, without breaking more symmetries than are already broken in $H_0$, we would obtain another description of the BFLS state.  However, translation symmetry requires both $t$- and $s$-fermions to be gapless at the $\Gamma$ point, and this state is thus distinct from the BFLS state, at least at the mean-field level.  It is conceivable that this mean-field state and the BFLS state could be two different free-fermion limits of the same phase.  Even if the two states are  distinct beyond mean-field theory, this state is quite similar to the BFLS state, and we expect that its physical properties are very similar to those elucidated in Ref.~\onlinecite{biswas11}.

\bibliography{msl}

\begin{thebibliography}{47}%
\makeatletter
\providecommand \@ifxundefined [1]{%
 \@ifx{#1\undefined}
}%
\providecommand \@ifnum [1]{%
 \ifnum #1\expandafter \@firstoftwo
 \else \expandafter \@secondoftwo
 \fi
}%
\providecommand \@ifx [1]{%
 \ifx #1\expandafter \@firstoftwo
 \else \expandafter \@secondoftwo
 \fi
}%
\providecommand \natexlab [1]{#1}%
\providecommand \enquote  [1]{``#1''}%
\providecommand \bibnamefont  [1]{#1}%
\providecommand \bibfnamefont [1]{#1}%
\providecommand \citenamefont [1]{#1}%
\providecommand \href@noop [0]{\@secondoftwo}%
\providecommand \href [0]{\begingroup \@sanitize@url \@href}%
\providecommand \@href[1]{\@@startlink{#1}\@@href}%
\providecommand \@@href[1]{\endgroup#1\@@endlink}%
\providecommand \@sanitize@url [0]{\catcode `\\12\catcode `\$12\catcode
  `\&12\catcode `\#12\catcode `\^12\catcode `\_12\catcode `\%12\relax}%
\providecommand \@@startlink[1]{}%
\providecommand \@@endlink[0]{}%
\providecommand \url  [0]{\begingroup\@sanitize@url \@url }%
\providecommand \@url [1]{\endgroup\@href {#1}{\urlprefix }}%
\providecommand \urlprefix  [0]{URL }%
\providecommand \Eprint [0]{\href }%
\providecommand \doibase [0]{http://dx.doi.org/}%
\providecommand \selectlanguage [0]{\@gobble}%
\providecommand \bibinfo  [0]{\@secondoftwo}%
\providecommand \bibfield  [0]{\@secondoftwo}%
\providecommand \translation [1]{[#1]}%
\providecommand \BibitemOpen [0]{}%
\providecommand \bibitemStop [0]{}%
\providecommand \bibitemNoStop [0]{.\EOS\space}%
\providecommand \EOS [0]{\spacefactor3000\relax}%
\providecommand \BibitemShut  [1]{\csname bibitem#1\endcsname}%
\let\auto@bib@innerbib\@empty
%</preamble>
\bibitem [{\citenamefont {Tsui}\ \emph {et~al.}(1982)\citenamefont {Tsui},
  \citenamefont {Stormer},\ and\ \citenamefont {Gossard}}]{tsui82}%
  \BibitemOpen
  \bibfield  {author} {\bibinfo {author} {\bibfnamefont {D.~C.}\ \bibnamefont
  {Tsui}}, \bibinfo {author} {\bibfnamefont {H.~L.}\ \bibnamefont {Stormer}}, \
  and\ \bibinfo {author} {\bibfnamefont {A.~C.}\ \bibnamefont {Gossard}},\
  }\href@noop {} {\bibfield  {journal} {\bibinfo  {journal} {Phys. Rev. Lett.}\
  }\textbf {\bibinfo {volume} {48}},\ \bibinfo {pages} {1559} (\bibinfo {year}
  {1982})}\BibitemShut {NoStop}%
\bibitem [{\citenamefont {Laughlin}(1983)}]{laughlin83}%
  \BibitemOpen
  \bibfield  {author} {\bibinfo {author} {\bibfnamefont {R.~B.}\ \bibnamefont
  {Laughlin}},\ }\href@noop {} {\bibfield  {journal} {\bibinfo  {journal}
  {Phys. Rev. Lett.}\ }\textbf {\bibinfo {volume} {50}},\ \bibinfo {pages}
  {1395} (\bibinfo {year} {1983})}\BibitemShut {NoStop}%
\bibitem [{\citenamefont {Wen}\ and\ \citenamefont {Niu}(1990)}]{wen90}%
  \BibitemOpen
  \bibfield  {author} {\bibinfo {author} {\bibfnamefont {X.~G.}\ \bibnamefont
  {Wen}}\ and\ \bibinfo {author} {\bibfnamefont {Q.}~\bibnamefont {Niu}},\
  }\href@noop {} {\bibfield  {journal} {\bibinfo  {journal} {Phys. Rev. B}\
  }\textbf {\bibinfo {volume} {41}},\ \bibinfo {pages} {9377} (\bibinfo {year}
  {1990})}\BibitemShut {NoStop}%
\bibitem [{\citenamefont {Anderson}(1987)}]{anderson87}%
  \BibitemOpen
  \bibfield  {author} {\bibinfo {author} {\bibfnamefont {P.~W.}\ \bibnamefont
  {Anderson}},\ }\href@noop {} {\bibfield  {journal} {\bibinfo  {journal}
  {Science}\ }\textbf {\bibinfo {volume} {235}},\ \bibinfo {pages} {1196}
  (\bibinfo {year} {1987})}\BibitemShut {NoStop}%
\bibitem [{\citenamefont {Sachdev}(2004)}]{sachdev04}%
  \BibitemOpen
  \bibfield  {author} {\bibinfo {author} {\bibfnamefont {S.}~\bibnamefont
  {Sachdev}},\ }in\ \href@noop {} {\emph {\bibinfo {booktitle} {Quantum
  Magnetism}}},\ \bibinfo {series} {Lecture Notes in Physics}, Vol.\ \bibinfo
  {volume} {645},\ \bibinfo {editor} {edited by\ \bibinfo {editor}
  {\bibfnamefont {U.}~\bibnamefont {Schollwock}}, \bibinfo {editor}
  {\bibfnamefont {J.}~\bibnamefont {Richter}}, \bibinfo {editor} {\bibfnamefont
  {D.~J.~J.}\ \bibnamefont {Farnell}}, \ and\ \bibinfo {editor} {\bibfnamefont
  {R.~F.}\ \bibnamefont {Bishop}}}\ (\bibinfo  {publisher} {Springer},\
  \bibinfo {year} {2004})\ pp.\ \bibinfo {pages} {381--432},\ \bibinfo {note}
  {(cond-mat/0401041)}\BibitemShut {NoStop}%
\bibitem [{\citenamefont {Senthil}(2004)}]{senthil04}%
  \BibitemOpen
  \bibfield  {author} {\bibinfo {author} {\bibfnamefont {T.}~\bibnamefont
  {Senthil}},\ }in\ \href@noop {} {\emph {\bibinfo {booktitle} {Recent Progress
  in Many-Body Theories: Proceedings of the 12th International Conference}}}\
  (\bibinfo  {publisher} {World Scientific},\ \bibinfo {year} {2004})\ \bibinfo
  {note} {(cond-mat/0411275)}\BibitemShut {NoStop}%
\bibitem [{\citenamefont {Alet}\ \emph {et~al.}(2006)\citenamefont {Alet},
  \citenamefont {Walczak},\ and\ \citenamefont {Fisher}}]{alet06}%
  \BibitemOpen
  \bibfield  {author} {\bibinfo {author} {\bibfnamefont {F.}~\bibnamefont
  {Alet}}, \bibinfo {author} {\bibfnamefont {A.~M.}\ \bibnamefont {Walczak}}, \
  and\ \bibinfo {author} {\bibfnamefont {M.~P.~A.}\ \bibnamefont {Fisher}},\
  }\href@noop {} {\bibfield  {journal} {\bibinfo  {journal} {Physica A}\
  }\textbf {\bibinfo {volume} {369}},\ \bibinfo {pages} {122} (\bibinfo {year}
  {2006})},\ \bibinfo {note} {(cond-mat/0511516)}\BibitemShut {NoStop}%
\bibitem [{\citenamefont {Lee}\ \emph {et~al.}(2006)\citenamefont {Lee},
  \citenamefont {Nagaosa},\ and\ \citenamefont {Wen}}]{lee06}%
  \BibitemOpen
  \bibfield  {author} {\bibinfo {author} {\bibfnamefont {P.~A.}\ \bibnamefont
  {Lee}}, \bibinfo {author} {\bibfnamefont {N.}~\bibnamefont {Nagaosa}}, \ and\
  \bibinfo {author} {\bibfnamefont {X.-G.}\ \bibnamefont {Wen}},\ }\href@noop
  {} {\bibfield  {journal} {\bibinfo  {journal} {Rev. Mod. Phys.}\ }\textbf
  {\bibinfo {volume} {78}},\ \bibinfo {pages} {17} (\bibinfo {year}
  {2006})}\BibitemShut {NoStop}%
\bibitem [{\citenamefont {Lee}(2008)}]{lee08}%
  \BibitemOpen
  \bibfield  {author} {\bibinfo {author} {\bibfnamefont {P.~A.}\ \bibnamefont
  {Lee}},\ }\href@noop {} {\bibfield  {journal} {\bibinfo  {journal} {Science}\
  }\textbf {\bibinfo {volume} {321}},\ \bibinfo {pages} {1306} (\bibinfo {year}
  {2008})}\BibitemShut {NoStop}%
\bibitem [{\citenamefont {Balents}(2010)}]{balents10}%
  \BibitemOpen
  \bibfield  {author} {\bibinfo {author} {\bibfnamefont {L.}~\bibnamefont
  {Balents}},\ }\href@noop {} {\bibfield  {journal} {\bibinfo  {journal}
  {Nature}\ }\textbf {\bibinfo {volume} {464}},\ \bibinfo {pages} {199}
  (\bibinfo {year} {2010})}\BibitemShut {NoStop}%
\bibitem [{\citenamefont {Wang}(2010)}]{wangf10}%
  \BibitemOpen
  \bibfield  {author} {\bibinfo {author} {\bibfnamefont {F.}~\bibnamefont
  {Wang}},\ }\href@noop {} {\bibfield  {journal} {\bibinfo  {journal} {Phys.
  Rev. B}\ }\textbf {\bibinfo {volume} {81}},\ \bibinfo {pages} {184416}
  (\bibinfo {year} {2010})}\BibitemShut {NoStop}%
\bibitem [{\citenamefont {Yao}\ and\ \citenamefont {Lee}(2011)}]{yao11}%
  \BibitemOpen
  \bibfield  {author} {\bibinfo {author} {\bibfnamefont {H.}~\bibnamefont
  {Yao}}\ and\ \bibinfo {author} {\bibfnamefont {D.~H.}\ \bibnamefont {Lee}},\
  }\href@noop {} {\bibfield  {journal} {\bibinfo  {journal} {Phys. Rev. Lett.}\
  }\textbf {\bibinfo {volume} {107}},\ \bibinfo {pages} {087205} (\bibinfo
  {year} {2011})}\BibitemShut {NoStop}%
\bibitem [{\citenamefont {Biswas}\ \emph {et~al.}(2011)\citenamefont {Biswas},
  \citenamefont {Fu}, \citenamefont {Laumann},\ and\ \citenamefont
  {Sachdev}}]{biswas11}%
  \BibitemOpen
  \bibfield  {author} {\bibinfo {author} {\bibfnamefont {R.~R.}\ \bibnamefont
  {Biswas}}, \bibinfo {author} {\bibfnamefont {L.}~\bibnamefont {Fu}}, \bibinfo
  {author} {\bibfnamefont {C.~R.}\ \bibnamefont {Laumann}}, \ and\ \bibinfo
  {author} {\bibfnamefont {S.}~\bibnamefont {Sachdev}},\ }\href@noop {}
  {\bibfield  {journal} {\bibinfo  {journal} {Phys. Rev. B}\ }\textbf {\bibinfo
  {volume} {83}},\ \bibinfo {pages} {245131} (\bibinfo {year}
  {2011})}\BibitemShut {NoStop}%
\bibitem [{\citenamefont {Lai}\ and\ \citenamefont {Motrunich}(2011)}]{lai11}%
  \BibitemOpen
  \bibfield  {author} {\bibinfo {author} {\bibfnamefont {H.-H.}\ \bibnamefont
  {Lai}}\ and\ \bibinfo {author} {\bibfnamefont {O.~I.}\ \bibnamefont
  {Motrunich}},\ }\href@noop {} {\bibfield  {journal} {\bibinfo  {journal}
  {Phys. Rev. B}\ }\textbf {\bibinfo {volume} {84}},\ \bibinfo {pages} {085141}
  (\bibinfo {year} {2011})}\BibitemShut {NoStop}%
\bibitem [{\citenamefont {Lai}\ and\ \citenamefont {Motrunich}()}]{lai11b}%
  \BibitemOpen
  \bibfield  {author} {\bibinfo {author} {\bibfnamefont {H.-H.}\ \bibnamefont
  {Lai}}\ and\ \bibinfo {author} {\bibfnamefont {O.~I.}\ \bibnamefont
  {Motrunich}},\ }\href@noop {} {}\Eprint {http://arxiv.org/abs/arXiv:1110.2581
  (2011)} {arXiv:1110.2581 (2011)} \BibitemShut {NoStop}%
\bibitem [{\citenamefont {Kitaev}(2006)}]{kitaev06}%
  \BibitemOpen
  \bibfield  {author} {\bibinfo {author} {\bibfnamefont {A.}~\bibnamefont
  {Kitaev}},\ }\href@noop {} {\bibfield  {journal} {\bibinfo  {journal} {Annals
  of Physics}\ }\textbf {\bibinfo {volume} {321}},\ \bibinfo {pages} {2}
  (\bibinfo {year} {2006})}\BibitemShut {NoStop}%
\bibitem [{\citenamefont {Greiter}\ and\ \citenamefont
  {Thomale}(2009)}]{greiter09}%
  \BibitemOpen
  \bibfield  {author} {\bibinfo {author} {\bibfnamefont {M.}~\bibnamefont
  {Greiter}}\ and\ \bibinfo {author} {\bibfnamefont {R.}~\bibnamefont
  {Thomale}},\ }\href@noop {} {\bibfield  {journal} {\bibinfo  {journal} {Phys.
  Rev. Lett.}\ }\textbf {\bibinfo {volume} {102}},\ \bibinfo {pages} {207203}
  (\bibinfo {year} {2009})}\BibitemShut {NoStop}%
\bibitem [{\citenamefont {Xu}\ and\ \citenamefont {Sachdev}(2010)}]{xu10a}%
  \BibitemOpen
  \bibfield  {author} {\bibinfo {author} {\bibfnamefont {C.}~\bibnamefont
  {Xu}}\ and\ \bibinfo {author} {\bibfnamefont {S.}~\bibnamefont {Sachdev}},\
  }\href@noop {} {\bibfield  {journal} {\bibinfo  {journal} {Phys. Rev. Lett.}\
  }\textbf {\bibinfo {volume} {105}},\ \bibinfo {pages} {057201} (\bibinfo
  {year} {2010})}\BibitemShut {NoStop}%
\bibitem [{\citenamefont {Wen}(2002)}]{wen02}%
  \BibitemOpen
  \bibfield  {author} {\bibinfo {author} {\bibfnamefont {X.-G.}\ \bibnamefont
  {Wen}},\ }\href@noop {} {\bibfield  {journal} {\bibinfo  {journal} {Phys.
  Rev. B}\ }\textbf {\bibinfo {volume} {65}},\ \bibinfo {pages} {165113}
  (\bibinfo {year} {2002})}\BibitemShut {NoStop}%
\bibitem [{\citenamefont {Alford}\ \emph {et~al.}(2008)\citenamefont {Alford},
  \citenamefont {Schmitt}, \citenamefont {Rajagopal},\ and\ \citenamefont
  {Sch\"afer}}]{alford08}%
  \BibitemOpen
  \bibfield  {author} {\bibinfo {author} {\bibfnamefont {M.~G.}\ \bibnamefont
  {Alford}}, \bibinfo {author} {\bibfnamefont {A.}~\bibnamefont {Schmitt}},
  \bibinfo {author} {\bibfnamefont {K.}~\bibnamefont {Rajagopal}}, \ and\
  \bibinfo {author} {\bibfnamefont {T.}~\bibnamefont {Sch\"afer}},\ }\href@noop
  {} {\bibfield  {journal} {\bibinfo  {journal} {Rev. Mod. Phys.}\ }\textbf
  {\bibinfo {volume} {80}},\ \bibinfo {pages} {1455} (\bibinfo {year}
  {2008})}\BibitemShut {NoStop}%
\bibitem [{\citenamefont {Zhou}\ and\ \citenamefont {Wen}()}]{zhou02}%
  \BibitemOpen
  \bibfield  {author} {\bibinfo {author} {\bibfnamefont {Y.}~\bibnamefont
  {Zhou}}\ and\ \bibinfo {author} {\bibfnamefont {X.~G.}\ \bibnamefont {Wen}},\
  }\href@noop {} {}\Eprint {http://arxiv.org/abs/cond-mat/0210662 (2002)}
  {cond-mat/0210662 (2002)} \BibitemShut {NoStop}%
\bibitem [{\citenamefont {Affleck}\ \emph {et~al.}(1988)\citenamefont
  {Affleck}, \citenamefont {Zou}, \citenamefont {Hsu},\ and\ \citenamefont
  {Anderson}}]{affleck88}%
  \BibitemOpen
  \bibfield  {author} {\bibinfo {author} {\bibfnamefont {I.}~\bibnamefont
  {Affleck}}, \bibinfo {author} {\bibfnamefont {Z.}~\bibnamefont {Zou}},
  \bibinfo {author} {\bibfnamefont {T.}~\bibnamefont {Hsu}}, \ and\ \bibinfo
  {author} {\bibfnamefont {P.~W.}\ \bibnamefont {Anderson}},\ }\href@noop {}
  {\bibfield  {journal} {\bibinfo  {journal} {Phys. Rev. B}\ }\textbf {\bibinfo
  {volume} {38}},\ \bibinfo {pages} {745} (\bibinfo {year} {1988})}\BibitemShut
  {NoStop}%
\bibitem [{\citenamefont {Dagotto}\ \emph {et~al.}(1988)\citenamefont
  {Dagotto}, \citenamefont {Fradkin},\ and\ \citenamefont {Moreo}}]{dagotto88}%
  \BibitemOpen
  \bibfield  {author} {\bibinfo {author} {\bibfnamefont {E.}~\bibnamefont
  {Dagotto}}, \bibinfo {author} {\bibfnamefont {E.}~\bibnamefont {Fradkin}}, \
  and\ \bibinfo {author} {\bibfnamefont {A.}~\bibnamefont {Moreo}},\
  }\href@noop {} {\bibfield  {journal} {\bibinfo  {journal} {Phys. Rev. B}\
  }\textbf {\bibinfo {volume} {38}},\ \bibinfo {pages} {2926} (\bibinfo {year}
  {1988})}\BibitemShut {NoStop}%
\bibitem [{Note1()}]{Note1}%
  \BibitemOpen
  \bibinfo {note} {Note that the projected wavefunction may in principle be
  magnetically ordered, or may break other symmetries besides spin rotation.
  One can check for this possibility by calculating the spin correlation
  function (or other appropriate correlation functions) after
  projection.}\BibitemShut {Stop}%
\bibitem [{\citenamefont {Wang}\ and\ \citenamefont
  {Vishwanath}(2006)}]{wangf06}%
  \BibitemOpen
  \bibfield  {author} {\bibinfo {author} {\bibfnamefont {F.}~\bibnamefont
  {Wang}}\ and\ \bibinfo {author} {\bibfnamefont {A.}~\bibnamefont
  {Vishwanath}},\ }\href@noop {} {\bibfield  {journal} {\bibinfo  {journal}
  {Phys. Rev. B}\ }\textbf {\bibinfo {volume} {74}},\ \bibinfo {pages} {174423}
  (\bibinfo {year} {2006})}\BibitemShut {NoStop}%
\bibitem [{\citenamefont {Senthil}\ and\ \citenamefont
  {Fisher}(2000)}]{senthil00}%
  \BibitemOpen
  \bibfield  {author} {\bibinfo {author} {\bibfnamefont {T.}~\bibnamefont
  {Senthil}}\ and\ \bibinfo {author} {\bibfnamefont {M.~P.~A.}\ \bibnamefont
  {Fisher}},\ }\href@noop {} {\bibfield  {journal} {\bibinfo  {journal} {Phys.
  Rev. B}\ }\textbf {\bibinfo {volume} {62}},\ \bibinfo {pages} {7850}
  (\bibinfo {year} {2000})}\BibitemShut {NoStop}%
\bibitem [{Note2()}]{Note2}%
  \BibitemOpen
  \bibinfo {note} {In the case $\protect \text {IGG} = {\protect \rm U}(1)$,
  the gauge field needs to be compact. This is so because for non-compact
  ${\protect \rm U}(1)$ gauge fields, the magnetic flux is a ${\protect \rm
  U}(1)$ globally conserved density, and therefore introducing a non-compact
  gauge field would introduce an extra global symmetry not present in the
  original spin model. This issue only arises when $\protect \text {IGG} =
  {\protect \rm U}(1)$.}\BibitemShut {Stop}%
\bibitem [{\citenamefont {Wen}(1991)}]{wen91}%
  \BibitemOpen
  \bibfield  {author} {\bibinfo {author} {\bibfnamefont {X.-G.}\ \bibnamefont
  {Wen}},\ }\href@noop {} {\bibfield  {journal} {\bibinfo  {journal} {Phys.
  Rev. B}\ }\textbf {\bibinfo {volume} {44}},\ \bibinfo {pages} {2664}
  (\bibinfo {year} {1991})}\BibitemShut {NoStop}%
\bibitem [{\citenamefont {Paramekanti}\ \emph {et~al.}(2004)\citenamefont
  {Paramekanti}, \citenamefont {Randeria},\ and\ \citenamefont
  {Trivedi}}]{paramekanti04}%
  \BibitemOpen
  \bibfield  {author} {\bibinfo {author} {\bibfnamefont {A.}~\bibnamefont
  {Paramekanti}}, \bibinfo {author} {\bibfnamefont {M.}~\bibnamefont
  {Randeria}}, \ and\ \bibinfo {author} {\bibfnamefont {N.}~\bibnamefont
  {Trivedi}},\ }\href@noop {} {\bibfield  {journal} {\bibinfo  {journal} {Phys.
  Rev. B}\ }\textbf {\bibinfo {volume} {70}},\ \bibinfo {pages} {054504}
  (\bibinfo {year} {2004})}\BibitemShut {NoStop}%
\bibitem [{\citenamefont {Hermele}\ \emph {et~al.}(2008)\citenamefont
  {Hermele}, \citenamefont {Ran}, \citenamefont {Lee},\ and\ \citenamefont
  {Wen}}]{hermele08}%
  \BibitemOpen
  \bibfield  {author} {\bibinfo {author} {\bibfnamefont {M.}~\bibnamefont
  {Hermele}}, \bibinfo {author} {\bibfnamefont {Y.}~\bibnamefont {Ran}},
  \bibinfo {author} {\bibfnamefont {P.~A.}\ \bibnamefont {Lee}}, \ and\
  \bibinfo {author} {\bibfnamefont {X.-G.}\ \bibnamefont {Wen}},\ }\href@noop
  {} {\bibfield  {journal} {\bibinfo  {journal} {Phys. Rev. B}\ }\textbf
  {\bibinfo {volume} {77}},\ \bibinfo {pages} {224413} (\bibinfo {year}
  {2008})}\BibitemShut {NoStop}%
\bibitem [{\citenamefont {Tay}\ and\ \citenamefont {Motrunich}(2011)}]{tay11}%
  \BibitemOpen
  \bibfield  {author} {\bibinfo {author} {\bibfnamefont {T.}~\bibnamefont
  {Tay}}\ and\ \bibinfo {author} {\bibfnamefont {O.~I.}\ \bibnamefont
  {Motrunich}},\ }\href@noop {} {\bibfield  {journal} {\bibinfo  {journal}
  {Phys. Rev. B}\ }\textbf {\bibinfo {volume} {83}},\ \bibinfo {pages} {235122}
  (\bibinfo {year} {2011})}\BibitemShut {NoStop}%
\bibitem [{\citenamefont {Ivanov}\ and\ \citenamefont
  {Senthil}(2002)}]{ivanov02}%
  \BibitemOpen
  \bibfield  {author} {\bibinfo {author} {\bibfnamefont {D.~A.}\ \bibnamefont
  {Ivanov}}\ and\ \bibinfo {author} {\bibfnamefont {T.}~\bibnamefont
  {Senthil}},\ }\href@noop {} {\bibfield  {journal} {\bibinfo  {journal} {Phys.
  Rev. B}\ }\textbf {\bibinfo {volume} {66}},\ \bibinfo {pages} {115111}
  (\bibinfo {year} {2002})}\BibitemShut {NoStop}%
\bibitem [{\citenamefont {Paramekanti}\ \emph {et~al.}(2005)\citenamefont
  {Paramekanti}, \citenamefont {Randeria},\ and\ \citenamefont
  {Trivedi}}]{paramekanti05}%
  \BibitemOpen
  \bibfield  {author} {\bibinfo {author} {\bibfnamefont {A.}~\bibnamefont
  {Paramekanti}}, \bibinfo {author} {\bibfnamefont {M.}~\bibnamefont
  {Randeria}}, \ and\ \bibinfo {author} {\bibfnamefont {N.}~\bibnamefont
  {Trivedi}},\ }\href@noop {} {\bibfield  {journal} {\bibinfo  {journal} {Phys.
  Rev. B}\ }\textbf {\bibinfo {volume} {71}},\ \bibinfo {pages} {094421}
  (\bibinfo {year} {2005})}\BibitemShut {NoStop}%
\bibitem [{\citenamefont {Burnell}\ and\ \citenamefont
  {Nayak}(2011)}]{burnell11}%
  \BibitemOpen
  \bibfield  {author} {\bibinfo {author} {\bibfnamefont {F.~J.}\ \bibnamefont
  {Burnell}}\ and\ \bibinfo {author} {\bibfnamefont {C.}~\bibnamefont
  {Nayak}},\ }\href@noop {} {\bibfield  {journal} {\bibinfo  {journal} {Phys.
  Rev. B}\ }\textbf {\bibinfo {volume} {84}},\ \bibinfo {pages} {125125}
  (\bibinfo {year} {2011})}\BibitemShut {NoStop}%
\bibitem [{\citenamefont {Senthil}\ and\ \citenamefont
  {Fisher}(2001)}]{senthil01}%
  \BibitemOpen
  \bibfield  {author} {\bibinfo {author} {\bibfnamefont {T.}~\bibnamefont
  {Senthil}}\ and\ \bibinfo {author} {\bibfnamefont {M.~P.~A.}\ \bibnamefont
  {Fisher}},\ }\href@noop {} {\bibfield  {journal} {\bibinfo  {journal} {Phys.
  Rev. B}\ }\textbf {\bibinfo {volume} {63}},\ \bibinfo {pages} {134521}
  (\bibinfo {year} {2001})}\BibitemShut {NoStop}%
\bibitem [{\citenamefont {Chen}\ \emph {et~al.}()\citenamefont {Chen},
  \citenamefont {Essin},\ and\ \citenamefont {Hermele}}]{unpublished_mft}%
  \BibitemOpen
  \bibfield  {author} {\bibinfo {author} {\bibfnamefont {G.}~\bibnamefont
  {Chen}}, \bibinfo {author} {\bibfnamefont {A.}~\bibnamefont {Essin}}, \ and\
  \bibinfo {author} {\bibfnamefont {M.}~\bibnamefont {Hermele}},\ }\href@noop
  {} {}\bibinfo {note} {~unpublished}\BibitemShut {NoStop}%
\bibitem [{\citenamefont {Xu}(2011)}]{xu10}%
  \BibitemOpen
  \bibfield  {author} {\bibinfo {author} {\bibfnamefont {C.}~\bibnamefont
  {Xu}},\ }\href@noop {} {\bibfield  {journal} {\bibinfo  {journal} {Phys. Rev.
  B}\ }\textbf {\bibinfo {volume} {83}},\ \bibinfo {pages} {024408} (\bibinfo
  {year} {2011})}\BibitemShut {NoStop}%
\bibitem [{\citenamefont {Hermele}(2007)}]{hermele07}%
  \BibitemOpen
  \bibfield  {author} {\bibinfo {author} {\bibfnamefont {M.}~\bibnamefont
  {Hermele}},\ }\href@noop {} {\bibfield  {journal} {\bibinfo  {journal} {Phys.
  Rev. B}\ }\textbf {\bibinfo {volume} {76}},\ \bibinfo {pages} {035125}
  (\bibinfo {year} {2007})}\BibitemShut {NoStop}%
\bibitem [{\citenamefont {Kim}(2006)}]{kskim06}%
  \BibitemOpen
  \bibfield  {author} {\bibinfo {author} {\bibfnamefont {K.-S.}\ \bibnamefont
  {Kim}},\ }\href@noop {} {\bibfield  {journal} {\bibinfo  {journal} {Phys.
  Rev. Lett.}\ }\textbf {\bibinfo {volume} {97}},\ \bibinfo {pages} {136402}
  (\bibinfo {year} {2006})}\BibitemShut {NoStop}%
\bibitem [{\citenamefont {Kim}(2007)}]{kskim07}%
  \BibitemOpen
  \bibfield  {author} {\bibinfo {author} {\bibfnamefont {K.-S.}\ \bibnamefont
  {Kim}},\ }\href@noop {} {\bibfield  {journal} {\bibinfo  {journal} {Phys.
  Rev. B}\ }\textbf {\bibinfo {volume} {75}},\ \bibinfo {pages} {245105}
  (\bibinfo {year} {2007})}\BibitemShut {NoStop}%
\bibitem [{\citenamefont {Yang}(1989)}]{cnyang89}%
  \BibitemOpen
  \bibfield  {author} {\bibinfo {author} {\bibfnamefont {C.~N.}\ \bibnamefont
  {Yang}},\ }\href@noop {} {\bibfield  {journal} {\bibinfo  {journal} {Phys.
  Rev. Lett.}\ }\textbf {\bibinfo {volume} {63}},\ \bibinfo {pages} {2144}
  (\bibinfo {year} {1989})}\BibitemShut {NoStop}%
\bibitem [{\citenamefont {Zhang}(1990)}]{zhang90}%
  \BibitemOpen
  \bibfield  {author} {\bibinfo {author} {\bibfnamefont {S.}~\bibnamefont
  {Zhang}},\ }\href@noop {} {\bibfield  {journal} {\bibinfo  {journal} {Phys.
  Rev. Lett.}\ }\textbf {\bibinfo {volume} {65}},\ \bibinfo {pages} {120}
  (\bibinfo {year} {1990})}\BibitemShut {NoStop}%
\bibitem [{\citenamefont {Tsvelik}(1992)}]{tsvelik92}%
  \BibitemOpen
  \bibfield  {author} {\bibinfo {author} {\bibfnamefont {A.~M.}\ \bibnamefont
  {Tsvelik}},\ }\href@noop {} {\bibfield  {journal} {\bibinfo  {journal} {Phys.
  Rev. Lett.}\ }\textbf {\bibinfo {volume} {69}},\ \bibinfo {pages} {2142}
  (\bibinfo {year} {1992})}\BibitemShut {NoStop}%
\bibitem [{\citenamefont {Coleman}\ \emph {et~al.}(1993)\citenamefont
  {Coleman}, \citenamefont {Miranda},\ and\ \citenamefont
  {Tsvelik}}]{coleman93}%
  \BibitemOpen
  \bibfield  {author} {\bibinfo {author} {\bibfnamefont {P.}~\bibnamefont
  {Coleman}}, \bibinfo {author} {\bibfnamefont {E.}~\bibnamefont {Miranda}}, \
  and\ \bibinfo {author} {\bibfnamefont {A.}~\bibnamefont {Tsvelik}},\
  }\href@noop {} {\bibfield  {journal} {\bibinfo  {journal} {Phys. Rev. Lett.}\
  }\textbf {\bibinfo {volume} {70}},\ \bibinfo {pages} {2960} (\bibinfo {year}
  {1993})}\BibitemShut {NoStop}%
\bibitem [{\citenamefont {Shastry}\ and\ \citenamefont
  {Sen}(1997)}]{shastry97}%
  \BibitemOpen
  \bibfield  {author} {\bibinfo {author} {\bibfnamefont {B.~S.}\ \bibnamefont
  {Shastry}}\ and\ \bibinfo {author} {\bibfnamefont {D.}~\bibnamefont {Sen}},\
  }\href@noop {} {\bibfield  {journal} {\bibinfo  {journal} {Phys. Rev. B}\
  }\textbf {\bibinfo {volume} {55}},\ \bibinfo {pages} {2988} (\bibinfo {year}
  {1997})}\BibitemShut {NoStop}%
\bibitem [{\citenamefont {Senthil}()}]{senthilunpub}%
  \BibitemOpen
  \bibfield  {author} {\bibinfo {author} {\bibfnamefont {T.}~\bibnamefont
  {Senthil}},\ }\href@noop {} {}\bibinfo {note} {~unpublished
  (2011)}\BibitemShut {NoStop}%
\bibitem [{\citenamefont {Sachdev}()}]{sachdevpc}%
  \BibitemOpen
  \bibfield  {author} {\bibinfo {author} {\bibfnamefont {S.}~\bibnamefont
  {Sachdev}},\ }\href@noop {} {}\bibinfo {note} {~private
  communication}\BibitemShut {NoStop}%
\end{thebibliography}%

\end{document}